\documentclass{article}
\usepackage{lmodern}

\usepackage{arxiv}

\usepackage[utf8]{inputenc} 
\usepackage[T1]{fontenc}    
\usepackage{hyperref}       
\usepackage{url}            
\usepackage{booktabs}       
\usepackage{amsfonts}       
\usepackage{nicefrac}       
\usepackage{microtype}      
\usepackage{lipsum}
\usepackage{graphicx}
\graphicspath{ {./images/} }
\usepackage{bm}
\newcommand{\mathsfbi}[1]{\bm{\mathit{#1}}}

\newcommand{\indexsize}{\small}
\usepackage[table]{xcolor}%
\usepackage{overpic}%
\usepackage{tikz}%
\usetikzlibrary{shapes.arrows}%
\usepackage[super]{nth}%
\usepackage{tikz}%
\usetikzlibrary{positioning}%
\usetikzlibrary{shapes.geometric}
\usepackage[super]{nth}
\usepackage{transparent}
\usetikzlibrary{positioning}
\usepackage{caption}

\usepackage{graphicx}
\usepackage{newtxtext}
\usepackage{newtxmath}
\usepackage[round]{natbib}
\usepackage{hyperref}
\hypersetup{
	linkcolor  = gray,
        colorlinks = true,
        urlcolor   = blue,
        citecolor  = black,
}

\newcommand{\RomanNumeralCaps}[1]
\linenumbers

\definecolor{BV_purple}{rgb}{0.54117, 0.1686, 0.8862}
\definecolor{blue_CFD}{rgb}{0.17647058823529413, 0.4588235294117647, 0.7137254901960784}
\definecolor{node_name}{rgb}{0.7451, 0.6549, 0.8980}
\definecolor{edge_name}{rgb}{0.44313, 0.44313, 0.44313}
\definecolor{Green}{rgb}{0, 0.5, 0}

\DeclareMathOperator*{\argmax}{arg\,max} 

\title{Discrete vortex-based broadcast mode analysis for mitigation of dynamic stall}

\author{
 Het D. Patel \\
  Department of Mechanical and Aerospace Engineering\\
  North Carolina State University\\
  Raleigh, NC 27606 \\
  \texttt{hdpatel3@ncsu.edu} \\
   \And
 Yi Tsung Lee \\
  Department of Mechanical and Aerospace Engineering\\
  North Carolina State University\\
  Raleigh, NC 27606 \\
  \texttt{lee26@ncsu.edu} \\
  \And
 Ashok Gopalarathnam\\
  Department of Mechanical and Aerospace Engineering\\
  North Carolina State University\\
  Raleigh, NC 27606 \\
  \texttt{agopalar@ncsu.edu} \\
  \And
  Chi-An Yeh\\
  Department of Mechanical and Aerospace Engineering\\
  North Carolina State University\\
  Raleigh, NC 27606 \\
  \texttt{chian.yeh@ncsu.edu} \\
 }

\begin{document}
\maketitle

\begin{abstract}

We integrate a discrete vortex method with complex network analysis to strategize dynamic stall mitigation over airfoils with active flow control. The objective is to inform the actuator placement and the timing to introduce control inputs during the highly transient process of dynamic stall. To this end, we treat a massively separated flow as a network of discrete vortical elements and quantify the interactions among the vortical nodes by tracking the spread of displacement perturbations between each pair of vortical elements using a discrete vortex method. This allows us to perform network broadcast mode analysis to identify an optimal set of discrete vortices, the critical timing, and the direction to seed perturbations as control inputs. Motivated by the objective of dynamic stall mitigation, the optimality is defined as maximizing the reduction of total circulation of the free vortices generated from the leading edge over a prescribed time horizon. We demonstrate the use of the analysis on a two-dimensional flow over a flat plate airfoil and a three-dimensional turbulent flow over a SD7003 airfoil. The results from the network analysis reveal that the optimal timing for introducing disturbances occurs slightly after the onset of flow separation, before the shear layer rolls up and forms the core of dynamic stall vortex. The broadcast modes also show that the vortical nodes along the shear layer are optimal for introducing disturbances, hence providing guidance to actuator placement. We validate these insights via flow simulations by placing an actuator near the leading edge of an airfoil to target the shear layer slightly after the separation onset. This results in a $21 \%$ and $14\%$ reduction in peak lift for flows over the flat plate and SD$7003$ airfoil, respectively. A corresponding decrease in vorticity injection from the airfoil surface under the influence of control is observed from simulations, which aligns with the objective of the network broadcast analysis. The study highlights the potential of integrating the discrete vortex methods with the network analysis to design an effective active flow control strategy for unsteady aerodynamics.

\end{abstract}

\section{Introduction}
\label{sec:Intro}

Dynamic stall is a critical flow phenomenon that significantly limits the performance of unsteady wings due to large-amplitude fluctuations in lift force beyond the static limit \citep[]{mccroskey1981phenomenon, carr1988progress}. This phenomenon is intrinsically linked to the unsteady separation of flow over the wing surface, culminating in the formation of a dynamic stall vortex (DSV) \citep{mccroskey1976dynamic}. The DSV is a dominant flow feature of dynamic stall flows, generated by the accumulation of surface vorticity flux concentrated near the leading edge during the wing's motion \citep{eldredge2019leading}. The severe unsteady aerodynamic loads associated with dynamic stall can impose significant performance and structural constraints, underscoring the importance of effective mitigation strategies with active flow control \citep[]{karim1994suppression, greenblatt2001dynamic, weaver2004control, kissing2021delaying}. 

While active flow control has demonstrated promising capabilities in mitigating dynamic stall, its design can be challenging due to the highly unsteady and transient nature of the flow \citep[]{gardner2017reduction, zhu2022combined, gardner2023review}. Particularly, the transient nature of the flow makes it difficult, yet even crucial, to determine an effective timing and location for control inputs to be introduced. The timing of actuation for dynamic stall flows is unclear due to the intricate interplay between the time scale of the flow dynamics and that of the wing motion \citep[]{feszty2004alleviation, tadjfar2018role, lombardi2013closed, de2025duty}. Such interplay also makes actuator placement a daunting question, since it causes separation point to travel over a great extent of the suction surface \citep{weaver2004control, singh2006control, post2006separation, kissing2020insights}. Therefore, there is a critical need for formal theoretic approaches to determine effective timing and location of actuation \citep[]{natarajan2016actuator, nair2021phase, godavarthi2023optimal} for mitigating dynamic stall.

Complex network analysis emerges as an appealing tool to address the need, since analogous questions arise in the field of network science \citep[]{Newman2010, albert2002statistical, Iacobello:review}. Determining a location for actuator placement is similar to identifying a hub for disease transmission over a human or animal network \citep{nande2021dynamics, safari2024modeling}. Modulating the timing of introducing control input is also analogous to determining when to implement preventive measures for disease outbreak, such as social distancing and isolation \citep{vinceti2020lockdown, schlosser2020covid}.  Therefore, not only does network science offer a broad and mature set of tools to address the aforementioned challenges for flow control \citep{yeh2021network, sujith2020complex}, but its use by the fluid dynamics community has also been witnessed in a growing number of studies, including model reduction \citep{kaiser2014cluster}, force prediction \citep{iacobello2022load}, and analyzing turbulent interactions \citep{iacobello2019lagrangian, tandon2023multilayer, li2021cluster}.

Constructing a network model requires the definitions of nodes and edge weights that quantify the interactions between the nodes. Establishing on a classic perspective in aerodynamics where fluid flows can be modeled as an ensemble of discrete vortices \citep{anderson2016, saffman1995vortex, eldredge2007numerical}, vortical elements have been adopted by many fluid dynamics studies as the nodes of fluid networks \citep{bai2019randomized, lander1993interaction, meena2021identifying}. To quantify the interactions between the vortical nodes, discrete vortex methods (DVMs) \citep{katz1981discrete, Koumoutsakos_vortex} become a natural choice due to their extensive applications to aerodynamic studies \citep{ramesh2014discrete, sureshbabu2019model, herndon2023linear, alvarez2020high}. In general, DVMs model a flow by time-evolving the locations of discrete vortices in a Lagrangian manner. Moreover, they are capable of modeling unsteady aerodynamic flows by introducing new vortical elements into the flow while determining their circulation strength upon formation \citep{Clements1973, sarpkaya1975inviscid, ramesh2014discrete}. Such introduction of finite-circulation vortices into the flow can be translated to vorticity injection that accounts for the formation of a dynamic stall vortex. Thus, in the context of DVMs, the objective of dynamic stall mitigation can be realized by modulating the strength of discrete vortices introduced to the flow, and such modulation can be achieved by perturbing the vortical elements present in the flow. Also, through an objective-informed combination of DVM and vortical network analysis, the effective timing and location of actuation can be identified by finding a critical set of vortical nodes at which introduced perturbation at an identified timing can result in a high level of modification in vorticity injection.

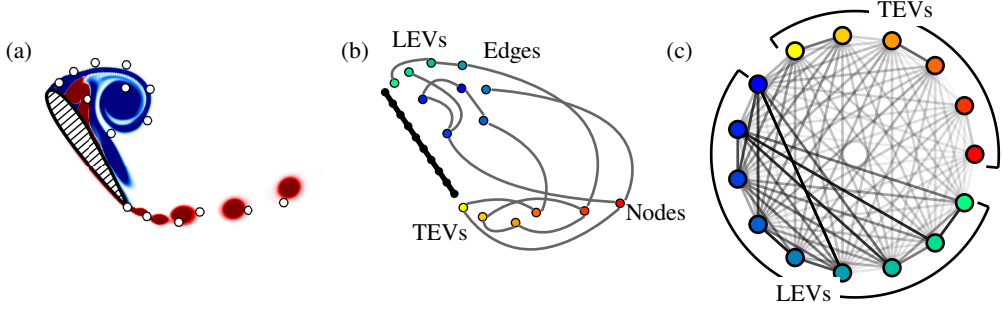
\begin{figure}
    \centering
    \vspace{-0.07in}
    \begin{overpic}[scale=1, trim = {0cm, 0cm, 0cm, 0cm}, clip]{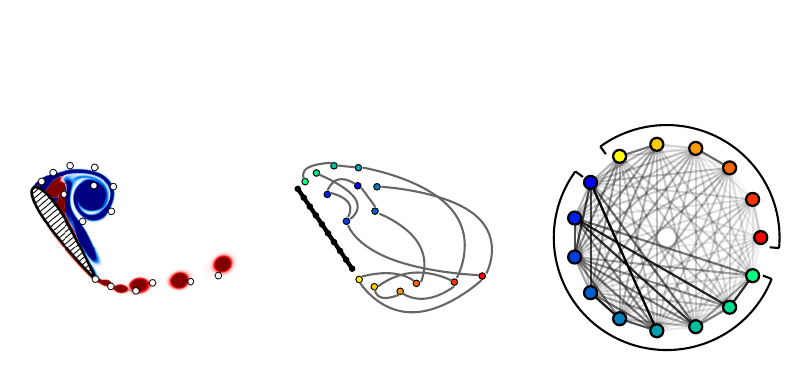} 
        \put(61, 11){\indexsize \textcolor{black}{Nodes}}
        \put(47, 27){\indexsize \textcolor{black}{Edges}}
        \put(00, 27){\indexsize \textcolor{black}{(a)}}
        \put(33, 27){\indexsize \textcolor{black}{(b)}}
        \put(65, 27){\indexsize \textcolor{black}{(c)}}

        \put(40, 09){\indexsize TEVs}
        \put(38, 28){\indexsize LEVs}

        \put(85, 31){\indexsize \colorbox{white}{TEVs}}
        \put(75, 03){\indexsize \colorbox{white}{LEVs}}
        
    \end{overpic}
    \caption{An overview of the present study: (a) a flow undergoing dynamic stall is modeled by a DVM \citep{ramesh2014discrete}; (b) the vortical elements are treated as network nodes to form a vortical network; (c) their interactions are extracted from the DVM as the edge weights to enable network analysis for mitigating dynamic stall. The vortices shown in (b) comprise of leading edge vortices (LEVs) and trailing edge vortices (TEVs).}
    \label{fig:methodolgy}
\end{figure}

In this study, we present a DVM-based network analysis to guide the timing, location, and direction of actuation for mitigating dynamic stall. To this end, we model the dynamic-stall flow as a time-varying network of discrete vortices and use a DVM \citep{lee2022state, ramesh2014discrete} to extract the interactions between the vortical nodes, as shown in figure \ref{fig:methodolgy}. Motivated by the objective of flow control, we quantify these vortical interactions (edge weights) by tracking the {\it perturbation spread} between the vortical elements with the use of the DVM. This edge-weight definition distinguishes the present study from many previous ones that adopted the induced velocity by the Biot--Savart law as edge weights of vortical networks \citep{nair2015network, de2023study, wang2025development}. Moreover, this integration of DVM and network analysis allows us to recast the questions of actuation timing, direction, and actuator location into an optimization problem that seeks an optimal set of vortical nodes to perturb and the timing to perturb them in a direction such that vorticity injection is minimized during dynamic stall and hence lift fluctuations can be mitigated.

In what follows, we discuss the problem setup of the dynamic stall flows in section \ref{sec:problem_setup}.  The construction of the DVM-based network model with the goal of guiding circulation control is elaborated in section \ref{sec:network_main}. In section \ref{sec:BR_ana}, we present the broadcast mode analysis \citep{grindrod2014, yeh2021network} to identify the optimal location and timing for actuation with the objective of dynamic stall mitigation by modulating vorticity injection. Section \ref{sec:result} leverages the insights from the analysis to inform flow control design and demonstrates the effectiveness of the network-based approach for mitigating dynamic stall.

\section{Problem setup} 
\label{sec:problem_setup}
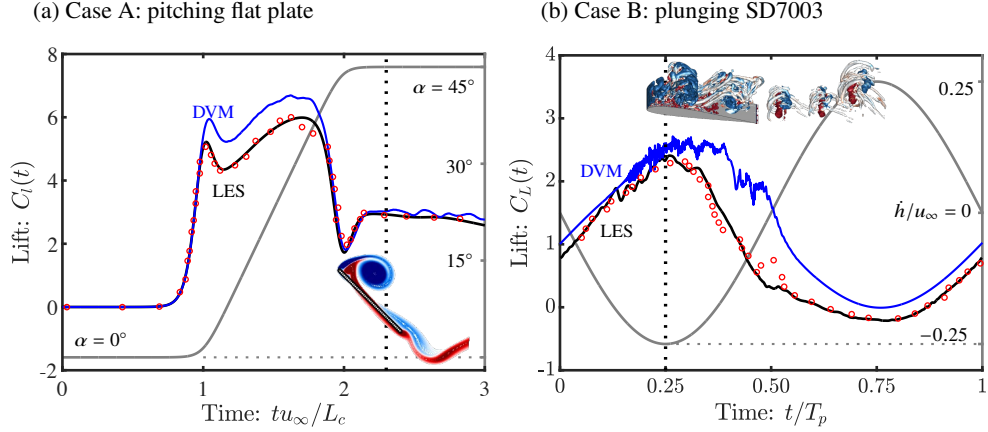
\begin{figure}
    \centering
    \vspace{0.2in}
    \begin{overpic}[scale=1]{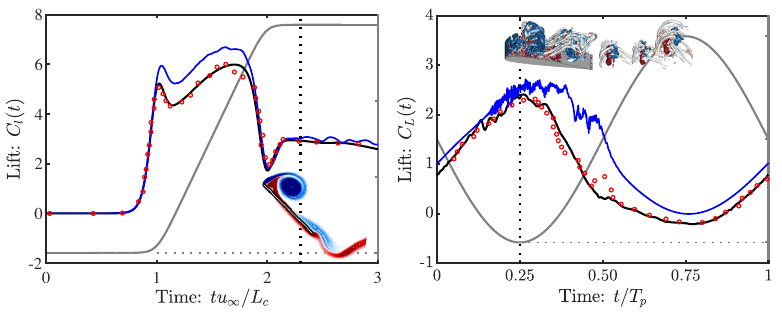} 
            \put(03, 42){\indexsize (a) Case A: pitching flat plate}
            \put(54, 42){\indexsize (b) Case B: plunging SD7003}


            \put(07, 09){\scriptsize $\alpha = 0^\circ$}
            \put(41.0, 34.5){\scriptsize $\alpha = 45^\circ$}
            \put(44.5, 26.4){\scriptsize $30^\circ$}
            \put(44.5, 17.0){\scriptsize $15^\circ$}
            \put(21, 24.0){\scriptsize LES}
            \put(19, 32.0){\scriptsize\color{blue} DVM}

            \put(94.0, 34.3){\scriptsize $0.25$}
            \put(89.5, 22){\scriptsize $\dot{h}/u_{\infty} = 0$}
            \put(92.0, 09.5){\scriptsize $-0.25$}
            \put(58, 26.0){\rotatebox{0}{\scriptsize\color{blue} DVM}}
            \put(60, 20.0){\rotatebox{0}{\scriptsize LES}}
    \end{overpic} 
    \caption{Two dynamic stall flows considered in this study: (a) a flow over a pitching flat plate airfoil (case A). The red circles ($\color{red}\circ$) are the results by \citet{ramesh2014discrete}; (b) flow over a periodic heaving-plunging SD$7003$ airfoil (case B). The red circles ($\color{red}\circ$) are the results by \citet{visbal2011numerical}. Contour lines of spanwise vorticity $\omega_{z}L_{c}/u_{\infty} \in [-20, 20]$ is shown for case (a). Iso-surface of $QL^{2}_{c}/u^{2}_{\infty} = 80$ colored by spanwise vorticity is shown for the case (b). Motion parameters in equation (\ref{eq:alpha_motion}) for case A: $\alpha_0 = \alpha_f = 22.5^\circ$, $aL_c/u_\infty = 11$, $b = 10.78$, $[t_1,~t_2]/(L_c/u_\infty) = [1,~1.98]$; motion parameters in equation (\ref{eq:plunge_motion}) for case B: $\nu_{0}/u_{\infty} = 0.25$, $T_{p}u_{\infty}/L_{c} = 4\pi$.}
    \label{fig:motion_flow}
\end{figure}

We consider two incompressible flows undergoing dynamic stall, as shown in figure \ref{fig:motion_flow}.  The first case (Case A) is a flow over a two-dimensional (2D) flat-plate airfoil in pitch-up motion at a chord-based Reynolds number, $Re \equiv u_{\infty}L_{c}/\nu_{\infty} = 1,000$. Here, $u_{\infty}$ is the free-stream velocity, $L_{c}$ is the chord length, and $\nu_{\infty}$ is the kinematic viscosity. For this case of pitching airfoil, the instantaneous angle of attack is prescribed via 
\begin{equation}
    {\alpha}(t) = \alpha_0 + \frac{\alpha_f}{b}
        \ln \left[
            \frac{\cosh\left(a (t - t_1)\right)}
                 {\cosh\left(a (t - t_2)\right)}\right],
    \label{eq:alpha_motion}
\end{equation}
following \citet{eldredge2009computational}. The second one (Case B) is a spanwise-periodic three-dimensional (3D) turbulent flow over a SD7003 airfoil undergoing periodic plunging-heaving motion at $Re = 60,000$, where the instantaneous vertical velocity of the airfoil is given by
\begin{equation} \label{eq:plunge_motion}
    \dot{h}(t) = -v_0 \sin\left(2\pi t / T_{p}\right),
\end{equation}
following \citet{visbal2011numerical, ramos2019active}. The motion parameters in equations (\ref{eq:alpha_motion}) and (\ref{eq:plunge_motion}) are provided in the caption of figure \ref{fig:motion_flow},  along with the motion profiles, instantaneous lift coefficients and flow visualizations at instants of DSV departure for both cases.

Baseline (uncontrolled) flows of both cases are obtained using large-eddy simulations (LES) and the discrete vortex method (DVM). The large-eddy simulations are performed using an incompressible flow solver, CharLES \citep{khalighi2011unstructured, bres2017unstructured}, on a body-fixed reference frame. This solver uses a node-based, second-order finite-volume method for spatial discretization and a fractional-step scheme for time stepping. To account for the unsteady motions of the airfoil, a time-dependent fictitious forcing is added to the right-hand-side of the momentum equation \citep{kim2006immersed}. A no-slip boundary condition is applied on the airfoil surface, and a time-varying free-stream condition is prescribed at the far-field boundary. For both the cases, the simulations are validated with respect to the lift fluctuations, for which the agreement is shown in figure \ref{fig:motion_flow}.

The DVM developed by \citet{lee2022state} and \cite{ramesh2014discrete} is employed in the present study.  It models unsteady flows over airfoils by introducing discrete point vortices into the flow, whose circulations are modulated by an empirical leading-edge suction parameter.  The time integration of the DVM adopts a fourth-order Runge-Kutta scheme, with a time step $\Delta t u_{\infty}/L_{c} = 0.01$.  The comparisons in lift obtained from the DVM and LES are also shown in figure \ref{fig:motion_flow}.  Although over-predictions in DVM lift are seen in both cases, we observe qualitative agreements in their temporal profiles.  Quantitatively, the levels of agreement in lift between DVM and CFD are also similar to those reported in previous studies \citep{ramesh2014discrete, lee2022state}.  The results of the DVM base flow will be used to construct the network model of discrete vortices, which is to be detailed in the following section.

\section{Construction of the discrete-vortex network} \label{sec:network_main}

Our goal of mitigating dynamic stall is approached by modeling the stall flows as networks of vortical elements, which enables a control-focused network analysis to provide guidance to effective actuation timing, location and direction. The first step of the network analysis is to construct an adjacency matrix, $\mathsfbi{A}$, with each of its entry, $A_{ij}$, representing a quantifiable interaction (edge weight) between a pair of vortical nodes $i$ and $j$. Below, we discuss how these vortical interactions are quantified with the use of a discrete vortex method (DVM).

\subsection{The discrete vortex method} \label{sec:DVM}
\begin{figure}
    \centering
    \begin{overpic}[scale = 1, trim = {0cm, 1.5cm, 0cm, 1.3cm}, clip]{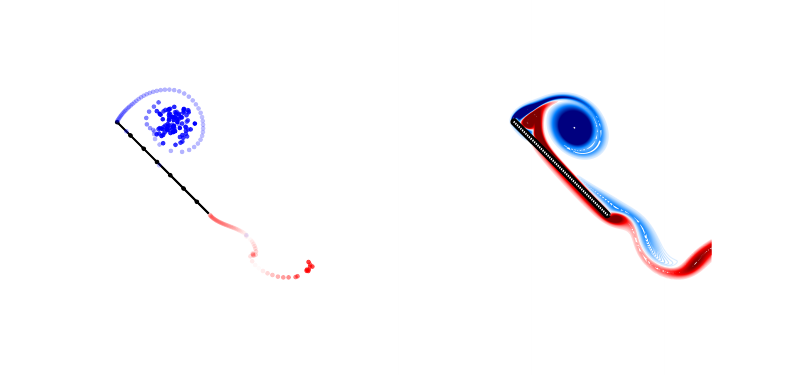}    
        \put(5, 25){(a)}
        \put(55, 25){(b)}
        
        \put(5, 14){\textcolor{black}{Bound vortices}}
        \put(25, -1){\textcolor{red}{trailing-edge vortices}}
        \put(25, 23){\textcolor{blue}{leading-edge vortices}}  

        \put(77, 20){\textcolor{blue}{Leading edge}}
        \put(77, 17){\textcolor{blue}{separated flow}}
        \put(80, -1){\textcolor{red}{wake flow}}
        \put(60, 15){\textcolor{black}{Airfoil}}
    \end{overpic}
    \caption{The flow state in DVM is shown in (a) with arrangement of bound, trailing-edge, and leading-edge vortical elements and its comparison with actual flow simulation is presented in (b).}
    \label{fig:DVM_intro}
\end{figure}
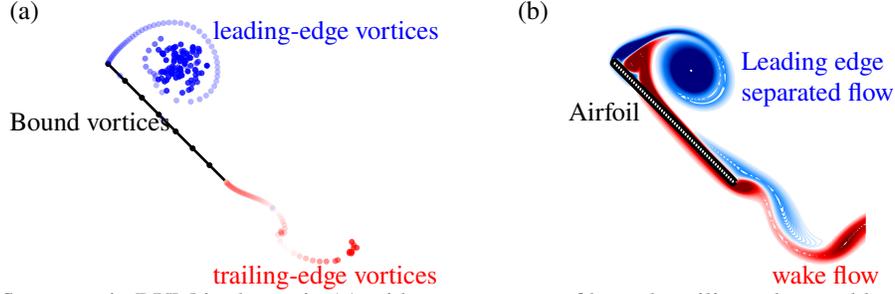

Here, we provide a concise overview on the DVM developed by \citet{ramesh2014discrete} and \citet{lee2022state}, which is employed in this study to quantify the interactions between the vortical nodes. Although some details below may be specific to the adopted DVM, our discussion will emphasize on how it facilitates the construction of the vortical network in a manner that the general concept is applicable to any DVMs \citep{Koumoutsakos_vortex, eldredge2007numerical}.  For more details about the DVM, the readers are referred to the original works by \citet{ramesh2014discrete} and \citet{lee2022state}.

The DVM simulates two-dimensional flows over an airfoil using three types of discrete vortices: leading edge vortices (LEVs), trailing edge vortices (TEVs), and bound vortices (BVs), as illustrated in figure \ref{fig:DVM_intro}. The BVs are distributed along the mean camber line of the airfoil and represent the airfoil location. The LEVs represent the flow separated from the leading edge of the airfoil, and the TEVs capture the flow past the trailing edge. Together, LEVs and TEVs are referred to as free vortices (FVs), since their locations vary in time. Here, we denote the locations and circulations of all FVs at time $t = t_{k}$ respectively by
\begin{equation*}
    \boldsymbol{\Xi}_{k}^{\text{fv}} 
    = 
    [
        \boldsymbol{\xi}_{1, k}^{\text{fv}},~
        \boldsymbol{\xi}_{2, k}^{\text{fv}},~\cdots,~ \boldsymbol{\xi}_{n_{\text{fv}}, k}^{\text{fv}}
    ] 
    \quad \text{and} \quad 
    \boldsymbol{\Gamma}^{\text{fv}}_{k} = 
    [
        \gamma^{\text{fv}}_{1},~
        \gamma^{\text{fv}}_{2},~\cdots,~ \gamma^{\text{fv}}_{n_{\text{fv}}}
    ],
\end{equation*}
where $n_{\text{fv}}$ is the number of FVs present in the flow at $t_k$. Note that $n_{\text{fv}}$ increases in time, i.e.~$n_{\text{fv}} = n_{\text{fv}}(t)$, since the DVM introduces new FVs to the flow at every time step.  Also, once a FV is introduced to the flow, its circulation is held constant in time so the time index, $k$, is not needed for $\gamma^{\text{fv}}$'s. Similarly, the circulations of the BVs at $t = t_{k}$ are denoted by
\begin{equation*}
    \boldsymbol{\Gamma}^{\text{bv}}_{k} 
    = 
    [
        \gamma^{\text{bv}}_{1, k},~
        \gamma^{\text{bv}}_{2, k},~\cdots,~ \gamma^{\text{bv}}_{n_{\text{bv}}, k}
    ],
\end{equation*}
where $n_{\text{bv}}$ is the number of BVs distributed along the airfoil. We also note that the locations of the BVs, $[\boldsymbol{\xi}^{\text{bv}}_{1, k},~\boldsymbol{\xi}^{\text{bv}}_{2, k},~\cdots,~\boldsymbol{\xi}^{\text{bv}}_{n_{\text{bv}}, k}]$, are given by a prescribed motion of the airfoil and hence are not the state variables to be determined by the DVM.

For each time step from $t_{k}$, the DVM seeds a TEV and a LEV respectively at the trailing edge ($\boldsymbol{\xi}^\text{tev}_k$) and leading edge ($\boldsymbol{\xi}^\text{lev}_k$)\footnote{In practice, the introduction of a LEV at a given instant is conditioned by the instantaneous leading-edge suction parameter \citep[]{ramesh2014discrete}. However, for an instant where a LEV is not introduced, this can be alternatively viewed as introducing a null LEV with zero circulation without loss of generality.}.  The circulations of the newly seeded TEV and LEV, respectively denoted as $\gamma^\text{tev}_{k+1}$ and $\gamma^\text{lev}_{k+1}$, are determined along with the circulations of all BVs, $\boldsymbol{\Gamma}^{\text{bv}}_{k+1}$, by imposing the no-penetration condition over the airfoil and the Kelvin's circulation theorem.  The new TEV and LEV are appended to the state vectors of FVs such that
$\boldsymbol{\Xi}^{\text{fv}}_{k'} = [
    \boldsymbol{\Xi}_{k}^{\text{fv}},~
    \boldsymbol{\xi}^{\text{tev}}_k,~
    \boldsymbol{\xi}^{\text{lev}}_k
]$ 
and 
$\boldsymbol{\Gamma}^{\text{fv}}_{k'} = 
	[
	\boldsymbol{\Gamma}^{\text{fv}}_{k},~
    \gamma^{\text{tev}}_{k+1},~
    \gamma^{\text{lev}}_{k+1}
	]
$, 
thereby increasing the number of FVs by 2. With the circulations for all BVs and new FVs determined, the DVM then updates the locations of all FVs in $\boldsymbol{\Xi}^{\text{fv}}_{k'}$ by
\[
    \boldsymbol{\xi}^{\text{fv}}_{i, k+1} 
    = \boldsymbol{\xi}^{\text{fv}}_{i, k} 
    + \Delta t \boldsymbol{\dot{\xi}}^{\text{fv}}_{i, k}, 
    \quad 
    \forall~i = 1,~2,~\dots,~n_{\text{fv}}(t_{k})+2,
\]
where $\Delta t$ is the time increment between $t_{k}$ and $t_{k+1}$, and $\boldsymbol{\dot{\xi}}^{\text{fv}}_{i, k}$ is the velocity of the $i$-th FV induced by all other discrete vortices. This induced velocity is given by
\[
\begin{split}
    \boldsymbol{\dot{\xi}}^{\text{fv}}_{i, k} 
    =
    \sum^{n_{\text{fv}}+2}_{j=1} \mathcal{V}_\text{bs}\left(\boldsymbol{\xi}^{\text{fv}}_{i, k},~\boldsymbol{\xi}^{\text{fv}}_{j, k},~\gamma^{\text{fv}}_{j}\right) 
    &+ \sum^{n_{\text{bv}}}_{j=1} \mathcal{V}_\text{bs}\left(\boldsymbol{\xi}^{\text{fv}}_{i, k},~ \boldsymbol{\xi}^{\text{bv}}_{j, k+1},~\gamma^{\text{bv}}_{j, k+1}\right),
\end{split}
\]
where $\mathcal{V}_\text{bs}$ is the Biot--Savart operator written as
\[
    \mathcal{V}_\text{bs}\left(\boldsymbol{\xi}_{i},~\boldsymbol{\xi}_{j},~\gamma_{j}\right) 
    = 
    \frac{\gamma_{j}}{2\pi} \frac{\boldsymbol{\hat{e}}_{z} \times (\boldsymbol{\xi}_{i} - \boldsymbol{\xi}_{j})}{|\boldsymbol{\xi}_{i} - \boldsymbol{\xi}_{j}|^{2} + r^{2}_{c}}, 
\]
with the core size $r_{c} = 1.3 \Delta t u_{\infty}$ \citep[]{lee2022state}. Once the locations of all FVs are updated, we obtain 
$
\boldsymbol{\Xi}^{\text{fv}}_{k+1} 
= 
	[
    \boldsymbol{\xi}^{\text{fv}}_{1, k+1},~
    \boldsymbol{\xi}^{\text{fv}}_{2, k+1},~
    \dots,~
    \boldsymbol{\xi}^{\text{fv}}_{n_{\text{fv}(t_k)}, k+1},~
    \boldsymbol{\xi}^{\text{tev}}_{k+1},~
    \boldsymbol{\xi}^{\text{lev}}_{k+1}
    ]
$, and the time iteration of the DVM is concluded with 
$
\boldsymbol{\Xi}^{\text{fv}}_{k+1}
$, 
$
	\boldsymbol{\Gamma}^{\text{fv}}_{k+1} = 
	\boldsymbol{\Gamma}^{\text{fv}}_{k'} = 
	[
	\boldsymbol{\Gamma}^{\text{fv}}_{k},~
    \gamma^{\text{tev}}_{k+1},~
    \gamma^{\text{lev}}_{k+1}
	]
$,
and
$\boldsymbol{\Gamma}^\text{bv}_{k+1}$. Summarizing the discussion above, we can express the time stepping operation of the DVM in brief as
\begin{equation} \label{eq:non_eq}
    \left[
        \boldsymbol{\Xi}^{\text{fv}}_{k+1},~\gamma^{\text{tev}}_{k+1},~\gamma^{\text{lev}}_{k+1},~ \boldsymbol{\Gamma}^{\text{bv}}_{k+1}
    \right] 
    = \mathcal{N} \left(\boldsymbol{\Xi}^{\text{fv}}_{k},~\boldsymbol{\Gamma}^{\text{fv}}_{k}\right),
\end{equation}
where $\mathcal{N}$ can be viewed as the discrete-time nonlinear state propagator of the DVM, whose nonlinearity stems from that in the Biot--Savart law.

Observing equation (\ref{eq:non_eq}), we make a few important notes in how the DVM can be integrated with network analysis to strategize dynamic stall mitigation:  first, equation (\ref{eq:non_eq}) suggests that the circulation introduced from the leading edge ($\gamma^{\text{lev}}_{k+1}$) can be modified by perturbing the locations of FVs ($\boldsymbol{\Xi}^{\text{fv}}_{k}$) present in the flow; second, if the location of a FV is perturbed at a time instant, the locations of all FVs will be perturbed at all future instants successively, resulting in continuous modification in the circulation introduced from the leading edge.  Since this circulation introduction can be translated to concentrated vorticity injection from the leading edge which results in the formation of DSV, equation (\ref{eq:non_eq}) also implies that dynamic stall mitigation can be potentially achieved by perturbing the FVs in the flow. Thus, the objective of the network analysis is to identify a set of FVs and an effective timing at which introduced perturbations to their locations can result in favorable modifications to the circulations of LEVs. Below, we discuss how this objective can be achieved by appropriately quantifying the interactions between the vortical elements.

\subsection{The time-evolving vortical network: adjacency and communicability matrices} \label{sec:network-DVM}

\begin{figure}
    \centering
    \begin{overpic}[width=1.0\textwidth]{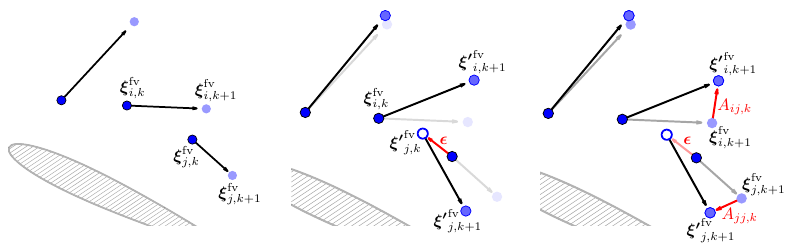}    
		\put(01, 29){\indexsize (a)}
		\put(37, 29){\indexsize (b)}
		\put(69, 29){\indexsize (c)}
    \end{overpic}
    \caption{Demonstration of the edge-weight calculation: (a) the translations of $i$-th and $j$-th FVs over a time step without perturbations; (b) An $\boldsymbol{\epsilon}$-perturbation is introduced to displace the $j$-th FV and changes the translations of $i$-th and $j$-th FVs over a time step; (c) the edge weight, $A_{ij, k}$, is obtained by calculating the differences between the perturbed and unperturbed translations for all FVs.}
    \label{fig:DVM_edgeweight}
\end{figure}

Motivated by the observations on equation (\ref{eq:non_eq}), we treat FVs in the DVM as the nodes in our network analysis. To quantify the interactions between them, we focus on those that take place due to displacement perturbations introduced to the vortical nodes, since such perturbations result in continuous change in the circulation introduced from the leading edge.

In this study, we construct the adjacency matrix by measuring the displacement perturbation passed between a pair of vortical nodes. Given a priori an instantaneous FV state at $t_{k}$ where $\boldsymbol{\Xi}^{\text{fv}}_{k}$ and $\boldsymbol{\Gamma}^{\text{fv}}_{k}$ are known, we define the edge weight, $A_{ij, k}$, as the resulting displacement of vortical node $i$ over a time increment due to a small displacement perturbation introduced to vortical node $j$. That is,
\begin{equation} \label{eq:edge_weight_elem}
    A_{ij, k} = \frac{\partial \boldsymbol{\xi}_{i, k+1}^{\text{fv}}}{\partial \boldsymbol{\xi}_{j, k}^{\text{fv}}}.
\end{equation}
Using the state propagator of the DVM (\ref{eq:non_eq}), this edge weight can be computed by
\begin{equation}
\label{eq:Ain_N}
    A_{ij, k} = \frac{1}{\epsilon} 
        \left.\left[\mathcal{N} \left(\boldsymbol{\Xi}^{\text{fv}}_{k} + \epsilon \boldsymbol{\Xi}^{'}(j),~\boldsymbol{\Gamma}^{\text{fv}}_{k}\right) 
        - 
        \mathcal{N} \left(\boldsymbol{\Xi}^{\text{fv}}_{k},~\boldsymbol{\Gamma}^{\text{fv}}_{k}\right)\right] \right\rvert_{\boldsymbol{\xi}_{i, k+1}^{\text{fv}}},
\end{equation}
where $\epsilon\boldsymbol{\Xi}^{'}(j)$ introduces an $\boldsymbol{\epsilon}$-perturbation to the location of $j$-th FV and zeros for all other FVs. The calculation for edge weights is demonstrated in figure \ref{fig:DVM_edgeweight}.  The translation of the $i$-th FV over a time step is changed from $\boldsymbol{\xi}^\text{fv}_{i, k+1}$ to $\boldsymbol{\xi'}^\text{fv}_{i, k+1}$ due to an $\boldsymbol{\epsilon}$-perturbation that displaces $\boldsymbol{\xi}^\text{fv}_{j, k}$ to $\boldsymbol{\xi'}^\text{fv}_{j, k}$, as shown in figure \ref{fig:DVM_edgeweight}b.  Thus, following equation (\ref{eq:Ain_N}), the edge weight, $A_{ij, k}$, is computed by
\[
    A_{ij, k} = \frac{1}{|\boldsymbol{\epsilon}|}\left(\boldsymbol{\xi'}^\text{fv}_{i, k+1} - \boldsymbol{\xi}^\text{fv}_{i, k+1}\right).
\]
The value of $|\boldsymbol{\epsilon}|$ is set to $10^{-7} r_{c}$, where linear convergence of each $A_{ij}$ is observed.  We note that, since both $\boldsymbol{\epsilon}$ and $\boldsymbol{\xi}$ are vectors, the resulting $A_{ij, k}$ is in fact a 2-by-2 matrix for the two-dimensional setting in the present study.  Here, we refer to the direction-embedded pairwise interaction as the edge weight for conceptual simplicity. Once the edge weights for all pairs of vortical nodes, $A_{ij, k}$, are computed, the adjacency matrix, $\mathsfbi{A}_{k} \in \mathbb{R}^{2n_\text{fv}\times2n_\text{fv}}$, can be fully constructed with a given $\boldsymbol{\Xi}^{\text{fv}}_{k}$ and $\boldsymbol{\Gamma}^{\text{fv}}_{k}$ at $t = t_{k}$.

Note that the locations of FVs, $\boldsymbol{\Xi}^{\text{fv}}_{k}$, vary in time. This results in a time-varying vortical network, where the edge weights and the constructed $\mathsfbi{A}_{k}$ become time dependent. In order to track the perturbation spread over such a network, a communicability matrix is required to account for the time-varying interactions between the vortical nodes \citep{grindrod2013matrix, grindrod2014}. The communicability matrix, $\mathsfbi{S}(t_{n}, t_{m})$, describes the interactions between the network nodes that occur over a finite time horizon from $t_{n}$ to $t_{m}$.   To construct $\mathsfbi{S}(t_{n}, t_{m})$, we first collect a series of snapshots for $\boldsymbol{\Xi}^{\text{fv}}_{k}$ and $\boldsymbol{\Gamma}^{\text{fv}}_{k}$ at each time step of DVM over the time interval $t \in [t_n, t_m]$.  These snapshots result in a series of adjacency matrices, $\mathsfbi{A}_{k}$'s, where $k$ is the time index of the FV state about which the adjacency matrix is constructed. Then, we follow \citet{Grindrod:PRE2011} and \citet{grindrod2013matrix} to construct the communicability matrix $\mathsfbi{S}(t_{n}, t_{m})$  by recurrently using
\begin{equation} \label{eq:Comm}
    \mathsfbi{S}(t_{n}, t_{k+1}) = \left[\boldsymbol{I} + \mathsfbi{A}_{k}\right]\left[\boldsymbol{I} + e^{-1}(\mathsfbi{S}(t_{n}, t_{k}) - \boldsymbol{I})\right],
\end{equation}
for $k = n,~n+1,~\dots,~m-1$, with $\mathsfbi{S}(t_{n}, t_{n}) = \boldsymbol{I}$, the identity matrix.  In the equation above, the time-downweighting factor, $e^{-1}$, enforces a finite-time decay of vortical perturbations and suppresses unbounded growth of their magnitudes.  This is conceptually similar to the exponential discounting in resolvent analysis \citep{Jovanovic:Thesis2004, Yeh:PRF2020} and can be motivated by the flow hysteresis during dynamic stall where the memory effect of wing motion eventually diminishes \citep{corke2015dynamic}.  For the present vortical perturbation network, each entry of the communicability matrix, or $S_{ij}$ of $\mathsfbi{S}(t_{n}, t_{m})$, represents the resulting displacement of the $i$-th vortical node at $t_{m}$ due to the displacement introduced at the $j$-th vortical node at an earlier time $t_{n}$.  That is, given a FV displacement perturbation at $t_n$, $\boldsymbol{\Xi}'_n$, the resulting displacement at $t_m$, $\boldsymbol{\Xi}'_m$, is related to $\boldsymbol{\Xi}'_n$ by $\mathsfbi{S}(t_{n}, t_{m})$ through
\begin{equation} \label{eq:Comm_st}
    \boldsymbol{\Xi}'_m = \mathsfbi{S}(t_n, t_m)\boldsymbol{\Xi}'_n.
\end{equation}
Thus, the communicability matrix, $\mathsfbi{S}(t_n, t_m)$, can be viewed as a state-transition operator of the displacement perturbation of FVs, $\boldsymbol{\Xi}'$, over a finite time horizon between $t_n$ and $t_m$.

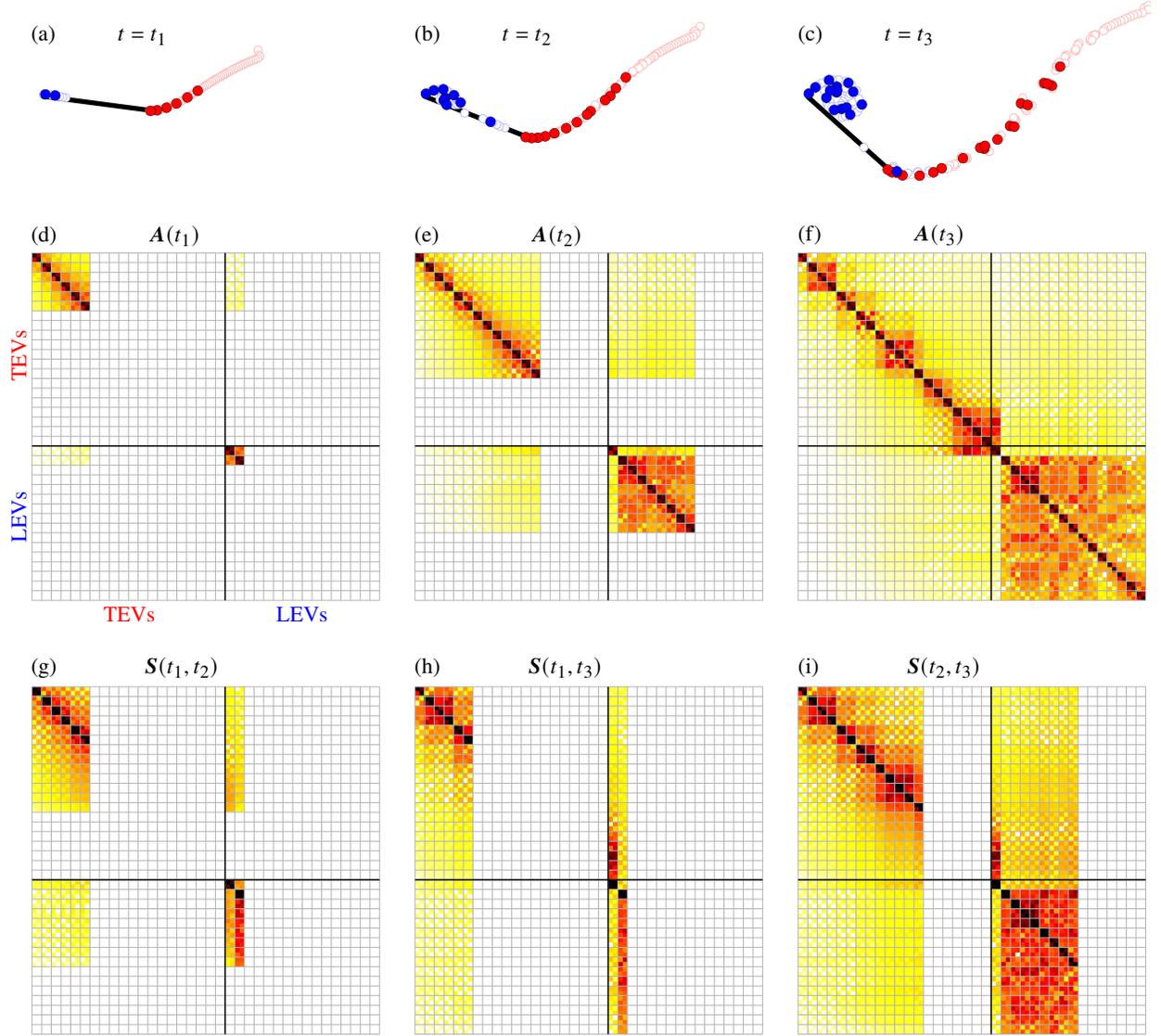
\begin{figure}
    \centering
    \begin{overpic}[width=1.0\textwidth]{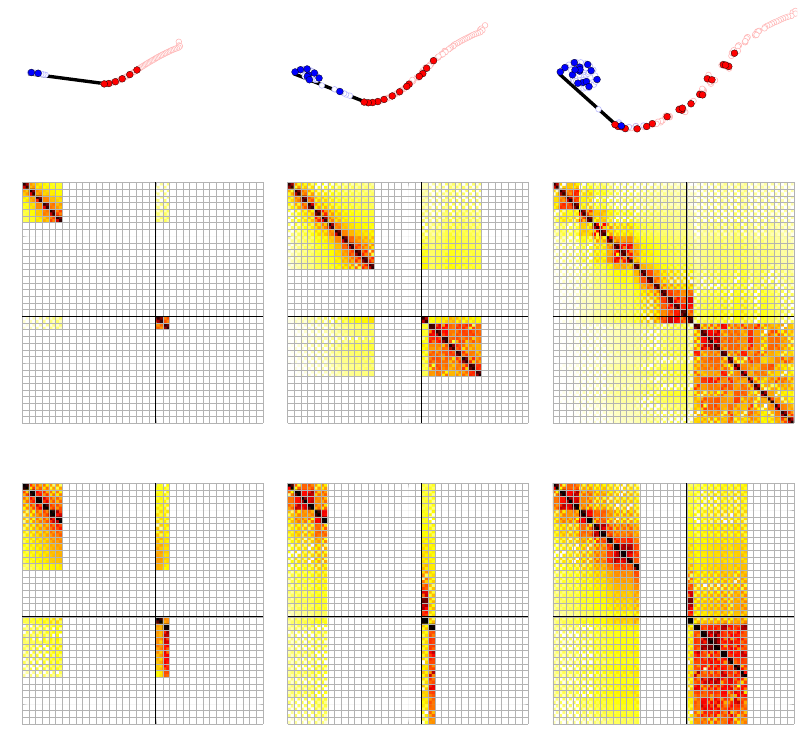}    
        \put(02.7, 87. ){\indexsize (a)\hspace{0.35in}$t = t_{1}$}
        \put(36.0, 87. ){\indexsize (b)\hspace{0.35in}$t = t_{2}$}
        \put(69.3, 87. ){\indexsize (c)\hspace{0.35in}$t = t_{3}$}
        
        \put(02.7, 69.5){\indexsize (d)\hspace{0.53in}$\mathsfbi{A}(t_{1})$}
        \put(36.0, 69.5){\indexsize (e)\hspace{0.53in}$\mathsfbi{A}(t_{2})$}
        \put(69.3, 69.6){\indexsize (f)\hspace{0.53in}$\mathsfbi{A}(t_{3})$}

        \put(02.7, 32.0){\indexsize (g)\hspace{0.5in}$\mathsfbi{S}(t_{1}, t_{2})$}
        \put(36.0, 32.0){\indexsize (h)\hspace{0.5in}$\mathsfbi{S}(t_{1}, t_{3})$}
        \put(69.3, 32.0){\indexsize (i)\hspace{0.5in}$\mathsfbi{S}(t_{2}, t_{3})$}
        
        \put(01, 59){\indexsize \rotatebox[]{90}{\textcolor{red}{TEVs}}}
        \put(01, 45){\indexsize \rotatebox[]{90}{\textcolor{blue}{LEVs}}}
        \put(09, 36.5){\indexsize \rotatebox[]{0}{\textcolor{red}{TEVs}}}
        \put(24, 36.5){\indexsize \rotatebox[]{0}{\textcolor{blue}{LEVs}}}
    \end{overpic}
    \caption{The adjacency and communicability matrices constructed about and between three representative time instances. (a-c) Locations of FVs at each instance; (d-f) Adjacency matrices constructed about each instance; (g-i) Communicability matrices constructed between the combinations of the instances.  For visual clarity, the matrices in (d-i) are sub-sampled such that only the interactions between the vortical elements highlighted by solid colors in (a-c) are shown.}
    \label{fig:temporal_mat}
\end{figure}

Let us examine in figure \ref{fig:temporal_mat} the adjacency and communicability matrices constructed at and between a few representative time instants, respectively. These are accompanied by the instantaneous $\boldsymbol{\Xi}^{\text{fv}}$ for the same set of instants. For the adjacency matrices at earlier times, we zero-pad the rows and columns corresponding to the vortical nodes that have yet been seeded into the flow. Also, each matrix is dissected into four blocks according to either TEVs or LEVs the rows and columns are associated with.  At the earliest instant $t_1$ where the LEV introduction is just initiated by the DVM, we observe high levels of intra-community interactions within the TEV and LEV groups from the diagonal blocks of $\mathsfbi{A}(t_{1})$.  Meanwhile, the off-diagonal blocks reveals weaker inter-community interactions between TEVs and LEVs, which is attributed to the longer distances between the TEVs and LEVs. At $t = t_2$, more LEVs and TEVs are introduced to the flow.  Their interactions populates the inactive entries of $\mathsfbi{A}(t_1)$ and emerge in the off-diagonal blocks of $\mathsfbi{A}(t_2)$.  Notably, while the interactions in the TEVs-to-TEVs block remain limited to the entries near the diagonal (close neighbors of vortical elements), we observe significantly richer interactions within LEV community, which can potentially be leveraged to mitigate dynamic stall.  Similar observations can be made for $\mathsfbi{A}(t_3)$, at which instant the DVM state is characterized by the formation of a DSV.  At this instant, we also observe that a LEV has advected to the trailing edge, which results in high levels of interactions between this LEV and the TEVs just seeded into the flow. 

Next, let us move our attention to the communicability matrices constructed between these three representative time instants.  Since each $S_{ij}$ element of $\mathsfbi{S}(t_{n}, t_{m})$ represents the displacement of the $i$-th vortex at $t_{m}$ due to a perturbation introduced to the $j$-th vortex at $t_{n}$, the $j$-th column of $\mathsfbi{S}(t_{n}, t_{m})$ represents the resulting displacements of all the vortices present at $t_{m}$ due to the perturbation of the $j$-th vortex at $t_{n}$.  Therefore, the columns of nonzero elements in $\mathsfbi{S}(t_{n}, t_{m})$ are the same as those in $\mathsfbi{A}(t_{n})$, and the rows of nonzero elements in $\mathsfbi{S}(t_{n}, t_{m})$ are the same as those in $\mathsfbi{A}(t_{m})$.  Such a structure of $\mathsfbi{S}(t_{n}, t_{m})$ suggests that only the vortical elements present at $t_n$ can be perturbed, and the perturbation can spread to all vortical elements present at $t_m$.  Comparing $\mathsfbi{S}(t_{1}, t_{2})$ with $\mathsfbi{A}(t_{1})$, we observe higher levels of interactions in the off-diagonal entries for both intra- and inter-community interactions within and between the TEV and LEV groups, respectively.  The same observations can be made by comparing $\mathsfbi{S}(t_{1}, t_{3})$ with $\mathsfbi{A}(t_{1})$ and $\mathsfbi{S}(t_{2}, t_{3})$ with $\mathsfbi{A}(t_{2})$, which is attributed to the spread of perturbations over time.  Similar to $\mathsfbi{A}$'s, stronger intra-community interactions within the LEV group are observed for $\mathsfbi{S}$'s.  Most importantly, all communicability matrices show stronger LEVs-to-TEVs interactions than TEVs-to-LEVs interactions.  This implies that perturbing LEVs is more effective in resulting in a global modification of discrete-vortex dynamics, compared to perturbing TEVs.

\subsection{The network broadcast analysis}\label{sec:BR_ana} \label{sec:BR_method}
Recall the remarks we made on the state propagator of the DVM (equation \ref{eq:non_eq}): it suggests that dynamic stall mitigation can potentially be achieved by identifying a set of FVs and an effective timing at which introduced perturbations result in desirable modifications to the LEV circulations.  This motivates the use of network broadcast analysis \citep{grindrod2014, yeh2021network} to identify the optimal timing and the corresponding set of FVs at which introduced perturbations effectively reduce the circulations of the LEVs.  In this analysis, the capability of each node in spreading information over a time-varying network is quantified by a {\it broadcast mode}, and that of each node in collecting information at a later time is distilled into a {\it receiving mode}.  Traditionally, the broadcast/receiving modes are obtained from either a column/row sum or a singular value decomposition of the communicability matrix \citep{grindrod2014, yeh2021network}.  Here, we customize its formulation by focusing on how introduced vortical perturbations result in desirable changes of the flow.  

In this study, we generalize the network broadcast analysis by formulating an optimization problem that seeks a set of vortical elements where introduced perturbations minimize the circulation generation from the leading edge over a specified time horizon.  Recall that the communicability matrix, $\mathsfbi{S} (t_n, t_k)$, serves as a finite-time state propagator for a displacement perturbation, $\boldsymbol{\Xi}'_n$, over $t \in [t_n, t_k]$, as discussed in equation (\ref{eq:Comm_st}).  The resulting displacement perturbation at $t_k$, or $\boldsymbol{\Xi}'_k$, will result in a change in circulation of the new LEV introduced at $t_{k+1}$.  Therefore, we can relate the change of LEV circulation at $t_{k+1}$, or $\Delta \gamma^\text{lev}_{k+1}$, to the displacement perturbation introduced at $t_n$ via a state-space representation, written as 
\begin{subequations}
\label{eq:state.space.gamma}
\begin{alignat}{2}
	\boldsymbol{\Xi}'_k &= \mathsfbi{S} (t_n, t_k) \boldsymbol{\Xi}'_n, \\
	\Delta \gamma^\text{lev}_{k+1} &= \sum_{i=1}^{n_\text{fv}(t_k)}\mathsfbi{C}_k  \boldsymbol{\Xi}'_k,
\end{alignat}
\end{subequations}
where $\mathsfbi{C}_{k}$ is a diagonal matrix in which each of its diagonal elements, 
\begin{equation}
	C_{ii, k} 
		= -\text{sgn}({\gamma^\text{lev}_{k+1}})\frac{\partial \gamma^\text{lev}_{k+1}}{\partial \boldsymbol{\xi}^\text{fv}_{i, k}},
\end{equation}
quantifies the reduction of LEV circulation due to a displacement perturbation introduced at the $i$-th vortical node at $t_k$.  Using the DVM state-propagator in (\ref{eq:non_eq}), we can compute these diagonal elements via
\begin{equation}
	C_{ii, k} =
		-\frac{\text{sgn}({\gamma^\text{lev}_{k+1}})}{\epsilon} 
        \left.\left[\mathcal{N} \left(\boldsymbol{\Xi}^{\text{fv}}_{k} + \epsilon \boldsymbol{\Xi}^{'}(i),~\boldsymbol{\Gamma}^{\text{fv}}_{k}\right) 
        - 
        \mathcal{N} \left(\boldsymbol{\Xi}^{\text{fv}}_{k},~\boldsymbol{\Gamma}^{\text{fv}}_{k}\right)\right] \right\rvert_{\gamma_{k+1}^{\text{lev}}}.
\end{equation}
Revisiting equation (\ref{eq:state.space.gamma}), now we can generalize the network broadcast analysis by solving an optimization problem \vspace{-0.1in}
\begin{equation} \label{eq:constraint}
\begin{aligned}
    \argmax_{\boldsymbol{\Xi}^\text{b}(t_n, t_m)} \quad\quad  
    \mathcal{J}(\boldsymbol{\Xi}^\text{b}) 
		&= \sum_{k = n}^m \Delta \gamma^\text{lev}_{k+1} 
		= \sum_{k = n}^m \sum_{i=1}^{n_\text{fv}(t_k)} \mathsfbi{C}_k \mathsfbi{S}(t_n, t_k) \boldsymbol{\Xi}^\text{b},\\
	\text{subject to} \quad\quad 
	{||{{{\boldsymbol{\Xi}^\text{b}}}}||}_2 &= 1,
\end{aligned}
\end{equation}
to seek a broadcast mode, $\boldsymbol{\Xi}^\text{b}(t_n, t_m)$, as an optimal displacement perturbation introduced to vortical elements at $t_n$ that maximizes the sum of the reduction in LEV circulations over all later times until $t_m$.  Meanwhile, the receiving mode, $\boldsymbol{\Xi}^\text{r}(t_n, t_m)$, is calculated via 
\begin{equation}
	\boldsymbol{\Xi}^\text{r}(t_n, t_m) = \mathsfbi{C}_k \mathsfbi{S}(t_n, t_m) \boldsymbol{\Xi}^\text{b},
\end{equation}
which quantifies the receptivity of each vortical element present at $t_m$ to the perturbation introduced at $t_n$ in the shape of the broadcast mode, $\boldsymbol{\Xi}^\text{b}$, with the objective of resulting in the maximum reduction in LEV circulations. 
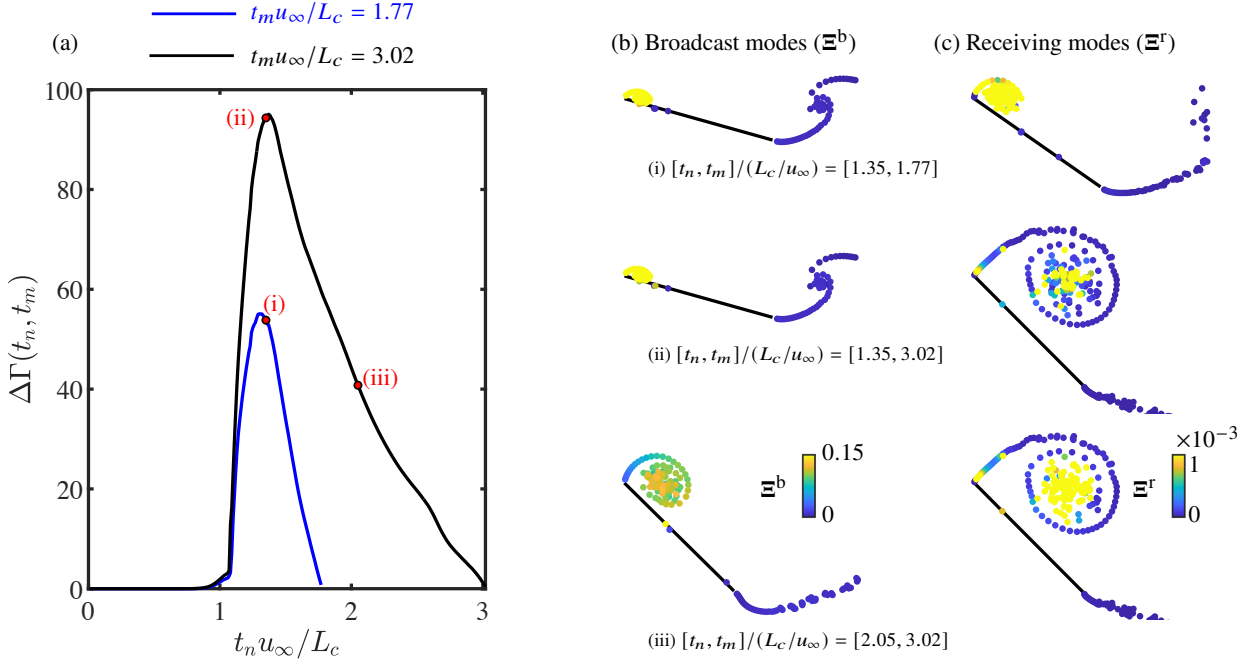
\begin{figure}
\vspace{0.2in}
    \centering
    \begin{overpic}[width=1.0\textwidth]{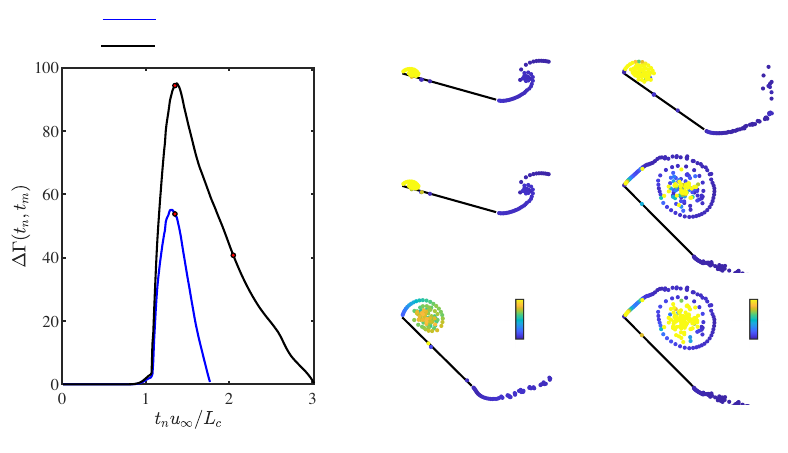}
        \put(05, 52){\indexsize (a)}
        \put(50, 52){\indexsize (b) Broadcast modes ($\boldsymbol{\Xi}^{\text{b}}$)}
        \put(76, 52){\indexsize (c) Receiving modes ($\boldsymbol{\Xi}^{\text{r}}$)}

        \put(53, 42){\scriptsize (i) $[t_n, t_m]/(L_c/u_\infty) = [1.35, 1.77]$}
        \put(53, 27){\scriptsize (ii) $[t_n, t_m]/(L_c/u_\infty) = [1.35, 3.02]$}
        \put(53, 04){\scriptsize (iii) $[t_n, t_m]/(L_c/u_\infty) = [2.05, 3.02]$}
        
        \put(22.0, 31.0){\indexsize\color{red} (i)}
        \put(19.0, 46.0){\indexsize\color{red} (ii)}
        \put(30.0, 25.0){\indexsize\color{red} (iii)}

        \put(62.0, 16.0){\indexsize $\boldsymbol{\Xi}^{\text{b}}$}
        \put(67.0, 14.0){\indexsize $0$}
        \put(67.0, 19.0){\indexsize $0.15$}
        
        \put(92.0, 16.0){\indexsize $\boldsymbol{\Xi}^{\text{r}}$}
        \put(96.5, 14.0){\indexsize $0$}
        \put(96.5, 18.0){\indexsize $1$}
        \put(95.5, 20.0){\indexsize $\times 10^{-3}$}

        \put(21.0, 54.5){\indexsize\color{blue} $t_{m}u_{\infty}/L_{c} = 1.77$}
        \put(21.0, 51.0){\indexsize\color{black} $t_{m}u_{\infty}/L_{c} = 3.02$}
    \end{overpic}
    \caption{A demonstration of the results of the broadcast mode analysis. The analysis yields three components: (a) the total reduction in LEV circulation between $t_n$ (perturbation time) and $t_m$ (a later time); (b) the broadcast modes $\boldsymbol{\Xi}^{\text{b}}$, which identify the optimal set of vortical nodes where introduced perturbations results in the highest reduction of LEV circulation; and (c) receiving modes $\boldsymbol{\Xi}^{\text{r}}$, which highlight the vortical nodes most influenced by the perturbations originating from the broadcast nodes. Three pairs of $[t_n, t_m]$ are chosen for the visualization of broadcast and response modes.}
    \label{fig:demo_BR}
\end{figure}
Note that the obtained broadcast mode, $\boldsymbol{\Xi}^\text{b}(t_n, t_m)$, is optimal for the chosen time horizon $t \in [t_n, t_m]$. And the resulting scalar objective function, $\mathcal{J}(\boldsymbol{\Xi}^\text{b}(t_n, t_m))$, corresponds to the total reduction in circulations of LEVs introduced to the flow over $t \in [t_n, t_m]$.  Thus, the optimal timing of introducing perturbations can be identified by the perturbation time $t_n$ that results in the maximum value of $\mathcal{J}(\boldsymbol{\Xi}^\text{b}(t_n, t_m))$ for a given $t_m$, which can be set to a critical instant associated with the DSV formation.  This is demonstrated in figure \ref{fig:demo_BR}, where the broadcast analysis is conducted between different combinations of $t_n$ and $t_m$, with $t_m > t_n$.  In particular, we sweep over different values of the perturbation time $t_n$ while keeping $t_m$ at two representative values, as shown in figure \ref{fig:demo_BR}a.  For three representative $t_n$-$t_m$ combinations, marked as (i), (ii), and (iii), the broadcast and receiving modes are visualized by coloring the FVs present at $t_n$ and $t_m$ respectively with the corresponding broadcast/receiving strengths, as quantified by the broadcast/receiving modes.  Note that the locations of the FVs of high broadcast strength will provide insights into the location of actuation.  Also, the perturbation time $t_n$ corresponding to high levels of $\Delta\Gamma(t_n, t_m) = \mathcal{J}(\boldsymbol{\Xi}^\text{b}(t_n, t_m))$ will be used to guide the timing of introducing control inputs for mitigating dynamic stall.  The use of the network broadcast analysis to guide active flow control will be discussed in detail in the next section.

\section{Result} \label{sec:result}
\subsection{Baseline flow physics} \label{sec:base_flow}
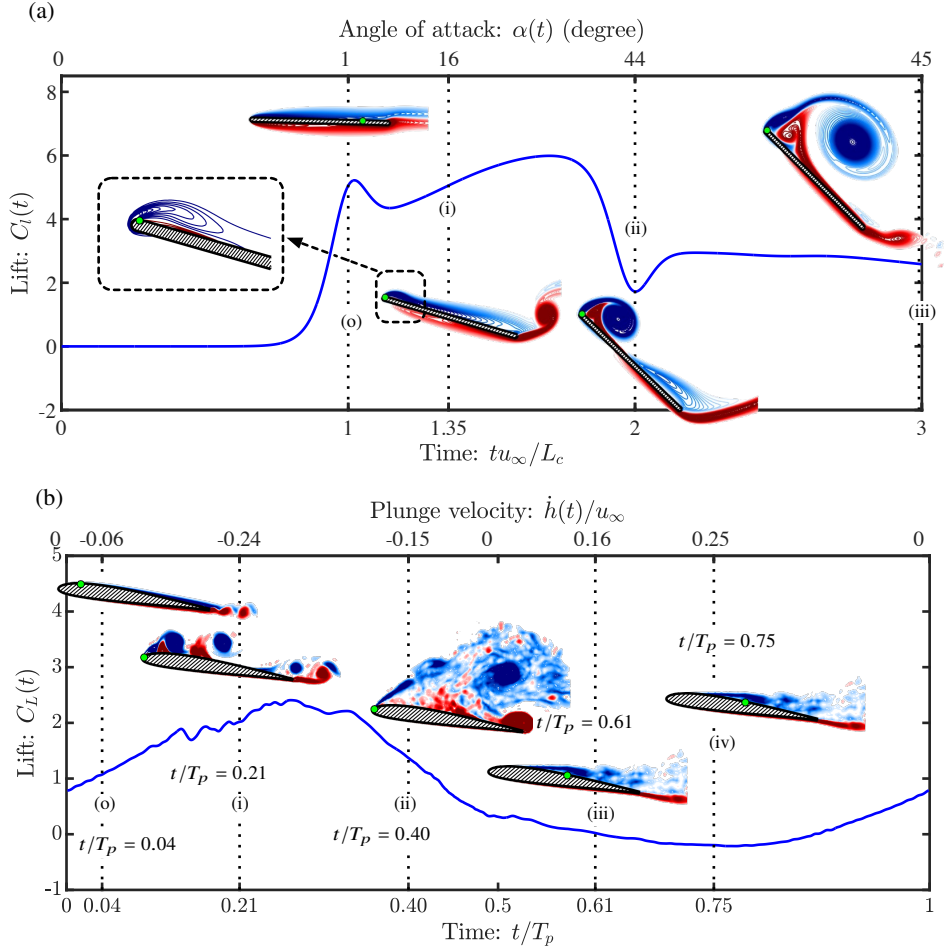
\begin{figure}
    \centering
    \begin{overpic}[scale=1]{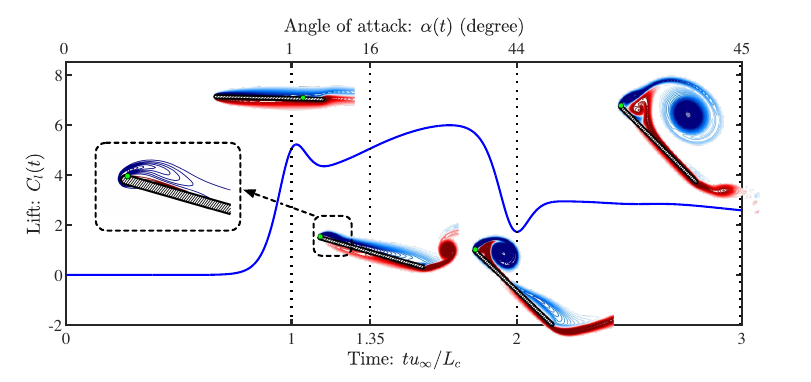} 
        \put (5, 45) {\indexsize (a)}
            
        \put (35, 15) {\setlength{\fboxsep}{1pt}\colorbox{white}{\scriptsize(o)}}
        \put (44.5, 26) {\setlength{\fboxsep}{1pt}\colorbox{white}{\scriptsize(i)}}
        \put (62.5, 24) {\setlength{\fboxsep}{1pt}\colorbox{white}{\scriptsize(ii)}}
        \put (90, 16) {\setlength{\fboxsep}{1pt}\colorbox{white}{\scriptsize(iii)}} 
    \end{overpic}
    
    \begin{overpic}[scale=1]{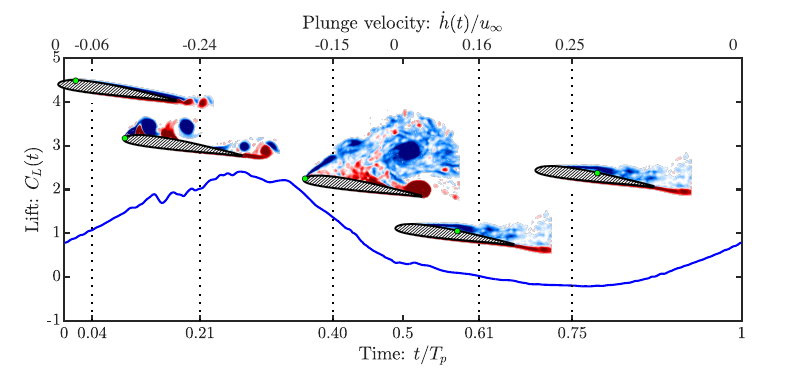} 
        \put (5, 45) {\indexsize (b)}

        \put (9, 11) {\setlength{\fboxsep}{1pt}\colorbox{white}{\scriptsize$t/T_{p} = 0.04$}}
        \put (18, 18) {\setlength{\fboxsep}{1pt}\colorbox{white}{\scriptsize$t/T_{p} = 0.21$}}
        \put (34, 12) {\setlength{\fboxsep}{1pt}\colorbox{white}{\scriptsize$t/T_{p} = 0.40$}}
        \put (54, 23) {\setlength{\fboxsep}{1pt}\colorbox{white}{\scriptsize$t/T_{p} = 0.61$}}
        \put (68, 31) {\setlength{\fboxsep}{1pt}\colorbox{white}{\scriptsize$t/T_{p} = 0.75$}}
        
        \put (10.5, 15) {\setlength{\fboxsep}{1pt}\colorbox{white}{\scriptsize (o)}}
        \put (24, 15) {\setlength{\fboxsep}{1pt}\colorbox{white}{\scriptsize (i)}}
        \put (40, 15) {\setlength{\fboxsep}{1pt}\colorbox{white}{\scriptsize (ii)}}
        \put (59, 14) {\setlength{\fboxsep}{1pt}\colorbox{white}{\scriptsize (iii)}}
        \put (71, 21) {\setlength{\fboxsep}{1pt}\colorbox{white}{\scriptsize (iv)}}
    
    \end{overpic}
    \caption{The lift fluctuations for (a) case A and (b) case B are shown alongside the spanwise vorticity fields at representative time instants. For case A, the close up view of the flow near the leading edge is presented in the dashed square for $tu_{\infty}/L_{c} = 1.35$. For case B, the vorticity fields are averaged in the spanwise direction. The separation point is marked with a green dot ($\color{green}\bullet$) in each vorticity field to identify leading edge separation.}
    \label{fig:cl_flow}
\end{figure}

To understand the dynamic stall flow physics of the two cases considered, the spanwise vorticity fields at a few representative instants are co-plotted along with the lift coefficients in figure \ref{fig:cl_flow}. The instantaneous angle of attack and the plunge velocity corresponding to the selected instants are marked along the top horizontal axis, respectively for case A (pitching flat plate) and case B (plunging SD7003). The instantaneous separation point at each instant is also mark by a green dot on the suction side to indicate the flow evolution to leading-edge separation as a signature of dynamic stall \citep{carr1977analysis, leishman1990dynamic}. 

In the beginning of pitching and downstroke motions, respectively for cases A and B, we observe that the separation point on the suction surface migrates upstream from the trailing edge, as indicated by instants (o) in figures~\ref{fig:cl_flow}a and \ref{fig:cl_flow}b. At a later instant (i), leading-edge separation occurs for both cases. At the same time, the stagnation point moves downstream along the pressure side of the airfoil. Consequently, the flow directed toward the suction surface from the stagnation point negotiates the leading edge curvature. This accelerates the flow near the leading edge, getting a strong suction force that contributes to rise in lift force. Moreover, at instant (i) negative vorticity accumulates over the suction surface near the leading edge, which marks the onset of DSV formation \citep{karim1994suppression}. Starting from instant (ii) for case A, the airfoil is held at $\alpha = 45^{\circ}$. The separation point remains near the leading edge, while the dynamic stall vortex convects away from the airfoil surface and results in decreasing lift, as indicated by instants (ii) and (iii) in figure~\ref{fig:cl_flow}a. For case B, the shear layer rolls up into smaller vortices due to Kelvin-Helmholtz instability \citep{visbal2011numerical, ramos2019active}. These vortices subsequently coalesce into the DSV, which is visible at instant (ii) in figure~\ref{fig:cl_flow}b. As the DSV departs from the airfoil surface, a reduction in the lift force from its peak value is observed. During the upstroke phase ($t/T_{p} \in [0.5, 1]$) for case B, the DSV passes beyond the trailing edge, accompanied by the shift of the separation point downstream along the suction surface. The flow gradually becomes fully attached, as seen at time instants (iii) and (iv) in figure~\ref{fig:cl_flow}b.

From the preceding discussion, it is evident that the formation of the DSV core is preceded by instants (i) in figures~\ref{fig:cl_flow}a and \ref{fig:cl_flow}b. This instant captures critical flow features that contribute to the formation of DSV core and eventually DSV in both cases. For case A, we characterize this instant by shear layer 'pinch-up', where the separated shear layer from the leading edge immediately reattaches to the suction surface at $x/L_{c} = 0.22$. At this instant, the accumulation of negative vorticity near the leading edge has initiated, which precedes the onset of shear layer roll-up that forms the DSV core. For the turbulent flow case B, the mechanism of DSV core formation differs from that of case A. Here, the shear layer roll-up into spanwise vortices has initiated in the high Reynolds number flow, as observed from instant (i) in figure~\ref{fig:cl_flow}b. Moreover, we observe that the formation of DSV core is driven by the merging of these spanwise vortices. This instant (i) for both cases marks the onset of DSV core formation and we will revisit them when discussing the results of the network analysis in the next section.

Since the dynamics of DSV plays a crucial role in the pressure distribution on the airfoil surface and lift fluctuation, interfering its formation process by modulating the vorticity injection from the leading edge can hold a key to mitigating dynamic stall. Next, we perform the broadcast mode analysis on the vortical network to identify the effective actuation timing and location, with the goal of achieving effective modulation of vorticity injection to mitigate dynamic stall.

\subsection{Broadcast mode analysis} \label{sec:outcome_BR}
\begin{figure}
    \centering
    \vspace{0.2in}
    \begin{overpic}[scale=1]{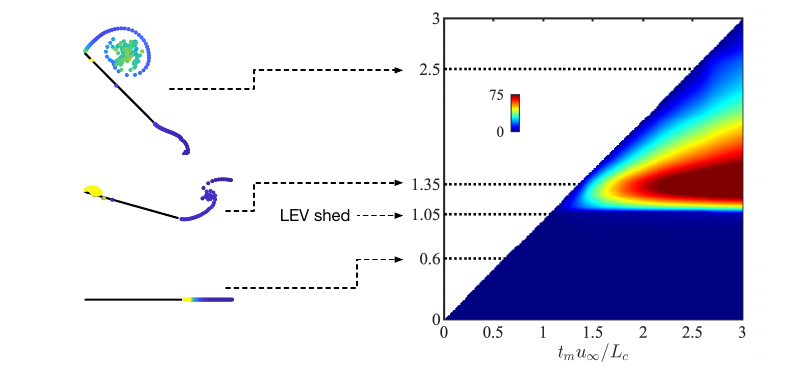}
        \put(00, 49){\indexsize (a)~~Case A: pitching flat plate}
        \put(52, 47){\indexsize $t_{n} u_{\infty}/L_c$}
        \put(02, 03){\indexsize Broadcast modes $\boldsymbol{\Xi}^\text{b}(t_n, t_m u_\infty/L_c = 3)$}
        \put(02, 11){\indexsize (o) $t_n u_\infty/L_c = 0.6$}
        \put(02, 18.5){\indexsize (i) $t_n u_\infty/L_c = 1.35$}
        \put(02, 29){\indexsize (ii) $t_n u_\infty/L_c = 2.5$}
        \put(66.5, 33){\indexsize $\Delta \Gamma$}
    \end{overpic}~\\
    \vspace{0.35in}
    \begin{overpic}[scale=1]{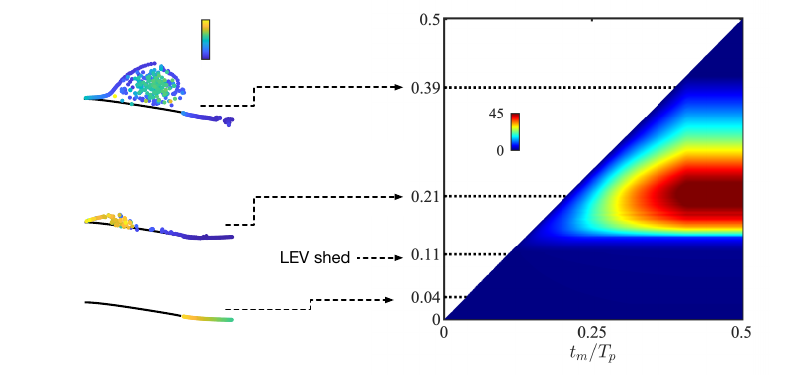}
        \put(00, 49){\indexsize (b)~~Case B: plunging SD7003}
        \put(52, 47){\indexsize $t_{n}/T_p$}
        \put(02, 03){\indexsize Broadcast modes $\boldsymbol{\Xi}^\text{b}(t_n, t_m /T_p = 0.5)$}
        \put(02, 10.5){\indexsize (o) $t_n/T_p = 0.04$}
        \put(02, 21.5){\indexsize (i) $t_n/T_p = 0.21$}
        \put(02, 31){\indexsize (ii) $t_n/T_p = 0.39$}
        \put(66.5, 30.5){\indexsize $\Delta \Gamma$}
        \put(22, 42){\indexsize $\boldsymbol{\Xi}^{\text{b}}$}
        \put(27, 40){\indexsize $0$}
        \put(27, 45){\indexsize $0.15$}
    \end{overpic}
\caption{The broadcast mode analyses for case A (a) and case B (b), sweeping over different combinations of perturbation time $t_n$ and a later time $t_m$.  For both cases, the reduction in LEV circulation, $\Delta \Gamma(t_{n}, t_{m})$, are shown on the right column, and the broadcast modes at three representative instances are visualized on the left column.  The broadcast modes are visualized by coloring the corresponding FV nodes by their broadcast strength.}
    \label{fig:gain_obj}
\end{figure}

\begin{figure}
    \centering
    \begin{overpic}[scale=1, trim={0cm, 0cm, 0cm, 0cm}, clip]{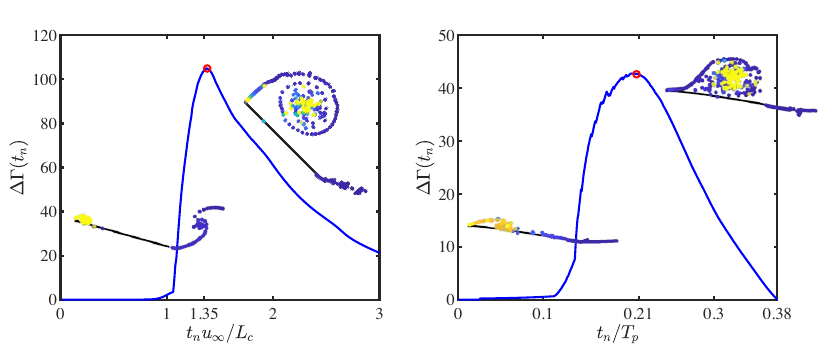}

        \put(8, 20){\indexsize Broadcast mode}
        \put(30, 36.5){\indexsize Receiving mode}

        \put(70, 12){\indexsize Broadcast mode}
        \put(80, 36.5){\indexsize Receiving mode}

        \put(01, 42){\indexsize (a) Case A: pitching flat plate}
        \put(52, 42){\indexsize (b) Case B: plunging SD$7003$}
    \end{overpic}
\caption{The profiles of $\Delta \Gamma(t_n, t_m)$ for a fix $t_m$ for (a) case A at $t_{m}u_{\infty}/L_{c} = 3$ and (b) case B at $t_{m}/T_{p} = 0.38$. The peak value of $\Delta \Gamma$ occurs at $t_{n}u_{\infty}/L_{c} = 1.35$ for case A and at $t_{n}/T_{p} = 0.21$ for case B, both marked by a red circle ($\color{red}\circ$). The corresponding broadcast and receiving modes are inserted for case A with $[t_{n}, t_{m}]/(L_{c}/u_{\infty})= [1.35, 3.00]$ and for case B with $[t_{n}, t_{m}]/T_{p} = [0.21, 0.38]$.}
\label{fig:gain_obj_line}
\end{figure}

As discussed in section~\ref{sec:BR_method}, the network broadcast mode analysis seeks a broadcast mode, $\boldsymbol{\Xi}^\text{b}(t_n, t_m)$, as an optimal displacement perturbation introduced to vortical elements at $t_n$ that maximizes the reduction in LEV circulations over all later times until $t_m$.  The objective function of the optimization problem is defined as the sum of the reduction in LEV circulations introduced to the flow over $t \in [t_n, t_m]$ due to a perturbation seeded at $t_n$. Therefore, the dependency of $\mathcal{J}(\boldsymbol{\Xi}^\text{b}(t_n, t_m)) = \Delta\Gamma(t_n, t_m)$ on the time of perturbation, $t_n$, and the resulting broadcast mode, $\boldsymbol{\Xi}^\text{b}(t_n, t_m)$, provide valuable insights into the effective timing and location for flow actuation.

This is shown in figure~\ref{fig:gain_obj}, where the broadcast mode analysis is conducted by sweeping over different combinations of perturbation time $t_n$ and a later time $t_m$. We obtain the contour plots for $\Delta\Gamma(t_n, t_m)$ below the main diagonal of figures~\ref{fig:gain_obj}a and \ref{fig:gain_obj}b, since $t_m > t_n$. Note that the introduction of LEVs in the DVM does not occur until $tu_\infty/L_c = 1.05$ for case A and $t/T_p = 0.11$ for case B. Consequently, the FV states at early times, as represented by instants (o) for both cases, are characterized by the presence of TEVs only. Hence, the broadcast modes at instants (o) are obtained only from these TEV nodes present in the flow. Moreover, the absence of LEVs at early times of perturbation ($t_n u_\infty/L_c < 1.05$ for case A and $t/T_p < 0.11$ for case B) leads to low levels of $\Delta\Gamma$ for both cases. This indicates that introducing perturbations at time instances characterized solely by the presence of TEVs is ineffective in modulating the circulation injected from the LE. This can be attributed to the convection of TEVs away from the airfoil surface that results in low levels of interactions with airfoil surface. 

Starting from the instants where LEV shedding is initiated, a rapid 
increase in $\Delta \Gamma$ is observed for both cases. The peaks of $\Delta \Gamma$ occurs at perturbation time $t_{n}u_{\infty}/L_{c} = 1.35$ for case A and $t_{n}/T_{p} = 0.21$ for case B, both after the initiation of LEV shedding. These peak instants are marked by instants (i) in figures~\ref{fig:gain_obj}a and \ref{fig:gain_obj}b. We make an important observation that these specific instants are corresponding to those leading to the DSV core formation, as previously discussed in figure~\ref{fig:cl_flow}. This suggests that the most effective timing of introducing perturbations for reducing LEV circulation corresponds to the early stages of DSV formation. Past instant (i), a decrease in $\Delta \Gamma$ is observed along with the concluded formation of DSV and its departure from the suction surface. This can be seen at the representative instants (ii) in figures~\ref{fig:gain_obj}a and \ref{fig:gain_obj}b for both cases. At these later time instants, while the DSV continues to accumulate circulations of weaker LEVs, the DSV has become a robust structure and begins to convect away from the airfoil surface. These observations underscore that once the DSV formation is concluded, the effectiveness of flow perturbation for modulating LEV circulation is significantly reduced.

Let us take a closer look at the $\Delta \Gamma(t_n,~t_m)$ profiles with respect to different perturbation time ($t_n$) and a fixed $t_m$ in figure~\ref{fig:gain_obj_line}.  Here, we choose $t_{m}u_{\infty}/L_{c} = 3$ for case A and $t_{m}/T_{p} = 0.38$ for case B and extract the $\Delta \Gamma$ values along the constant-$t_m$ lines from figure \ref{fig:gain_obj}. Note that both of the chosen instants are associated with the concluded formation of DSV over the suction surface. The resulting trends of $\Delta \Gamma(t_n)$ are shown in figures~\ref{fig:gain_obj_line}a and \ref{fig:gain_obj_line}b for case A and case B, respectively.  In particular, the broadcast and receiving modes obtained from the peak $t_n$, highlighted by a red dot on the $\Delta \Gamma(t_n)$ profile, are also inserted for both cases.  Once again, we observe that the broadcast modes, $\boldsymbol{\Xi}^{\text{b}}$, highlight the vortical elements residing along the shear layer emanating from the LE. This suggests that perturbations introduced to the shear layer can potentially induce high levels of LEV circulation reduction, as quantified by $\Delta\Gamma$. Moreover, the receiving modes, $\boldsymbol{\Xi}^{\text{r}}$, obtained at the chosen later times $t_{m}$, reveal that the LEVs residing within the DSV core are those perturbed the most by the perturbations seeded into the flow at the earlier perturbation time $t_{n}$ that results in the highest $\Delta\Gamma$. These observations indicate that the optimal timing in introducing perturbations for active flow control occurs slightly after the onset of LE separation and prior to the formation of a DSV core, as identified by the LEV shedding in the DVM. With the identified timing for effective actuation, next we leverage the insights from the broadcast modes to identify an effective location on the airfoil surface for actuator placement. 
 
\begin{figure}
    \centering
    \begin{overpic}[scale = 1, trim = {0cm, 0cm, 0cm, 0cm}, clip]{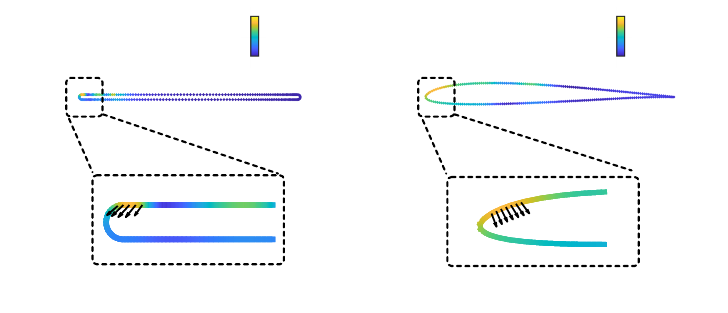}        
        \put(-3, 41){\indexsize (a) Case A: pitching flat plate}
        \put(49, 41){\indexsize (b) Case B: plunging SD$7003$}

        \put(36, 35){\indexsize $0$}
        \put(36, 40){\indexsize $1$}
        \put(30, 32.5){\indexsize $\lvert \Delta \boldsymbol{u} \rvert/|\Delta \boldsymbol{u}|_\infty$}

        \put(86.5, 35){\indexsize $0$}
        \put(86.5, 40){\indexsize $1$}
        \put(81, 32.5){\indexsize $\lvert \Delta \boldsymbol{u} \rvert/|\Delta \boldsymbol{u}|_\infty$}

    \end{overpic}
    \vspace{-0.2in}
    \caption{The induced velocity on the airfoil surface due to the broadcast-mode perturbation. The close-up view of the airfoil section with high magnitude of perturbation velocity, $\lvert \Delta \boldsymbol{u}\rvert$, is presented along with its direction.}
    \label{fig:BR_direction}
\end{figure}
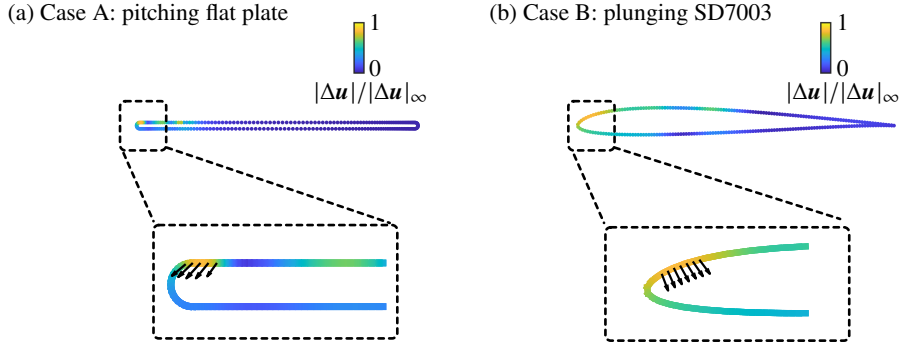

Note that the broadcast mode represents the optimal displacement perturbation applied to the Lagrangian vortical nodes. We can translate this Lagrangian perturbation to an Eulerian one by computing the induced velocity perturbation at an Eulerian grid point due to a displacement perturbation introduced at a Lagrangian vortical node by the Biot-Savart operator, $\mathcal{V}_\text{bs}$, used in the DVM. In this study, we aim to place an actuator on the surface of the airfoil. To this end, we compute the velocity perturbation on the airfoil surface due to a broadcast-mode perturbation via
\begin{equation} \label{eq:perturb_vel_field}
        \Delta \boldsymbol{u}(\boldsymbol{x}_{s}, k) = \sum^{n_{\text{fv}}}_{j = 1} \left[\mathcal{V}_{\text{bs}} \left(~\boldsymbol{x}_{s},~\boldsymbol{\xi}^{\text{fv}}_{j, k} + \epsilon \boldsymbol{\Xi}_{j, k}^{\text{b}},~\gamma^{\text{fv}}_{j}\right)
    -    \mathcal{V}_{\text{bs}}\left(\boldsymbol{x}_{s},~\boldsymbol{\xi}^{\text{fv}}_{j, k},~\gamma^{\text{fv}}_{j}\right)\right],
\end{equation}
where $\boldsymbol{x}_{s}$ parametrizes a set of Eulerian points on the airfoil surface at which the velocity perturbation is sought, and the time index $k$ is associated with the time instants where $\Delta \Gamma$ reaches its maximum. This will allow us to choose the actuator locations according to the magnitude of the velocity perturbation, $\lvert \Delta \boldsymbol{u} \rvert$. This is shown in figure~\ref{fig:BR_direction}, where we observe a region near the LE exhibits high levels of $\lvert \Delta \boldsymbol{u} \rvert$ for both cases. This occurs around $x/L_{c} = 0.02$ from the pitching flat plate (case A) and $x/L_{c} = 0.04$ for the plunging SD7003 (case B), indicating potentially effective locations to introduce control inputs. Moreover, when momentum-based actuators \citep{cattafesta2011actuators} are deployed for flow control, we can also inform the direction of actuation according to the orientation of the velocity perturbation at the peak location. These directions for both cases are also inserted in figures~\ref{fig:BR_direction}a and \ref{fig:BR_direction}b for case A and case B, respectively. Note that the optimal direction shows a combination of both wall-normal and tangential components for both cases.

As a short summary, the broadcast mode analysis provided us with insights into the effective timing, location and direction for flow actuation. It suggests that applying actuation near the leading edge shortly before the formation of DSV core can favorably reduce vorticity injection, potentially mitigating dynamic stall. Next, these insights are leveraged to inform flow control design, and their control effectiveness are examined via large-eddy simulations of controlled flows.

\subsection{Broadcast mode analysis guided active flow control } \label{sec:result_section}
In this section, we design active flow control based on the results of broadcast mode analysis and examine its effectiveness in mitigating dynamic stall using LES of controlled flows. Here, we employ a flow actuator \citep{cattafesta2011actuators} on the airfoil surface, according to the insights obtained from figure \ref{fig:BR_direction}.
This actuator is implemented as a time-varying Dirichlet boundary condition on the airfoil surface, where the velocity profile is given by
\begin{equation} \label{eq:surf_vel}
    u_\text{act} (\boldsymbol{x}_s, t) = u_\text{suction} \times g(\boldsymbol{x}_s) \times \phi(t).
\end{equation}
Here, $u_\text{suction}$ controls the suction strength, $g(\boldsymbol{x}_s)$ provides a compact spatial support for the actuator on the airfoil surface, and $\phi(t)$ controls the temporal duty cycle of the actuation input. The specific forms and  parameters for $g(\boldsymbol{x}_s)$ and $\phi(t)$ are provided in table \ref{tab:kd}. In particular, the temporal profile is chosen such that the suction strength reaches $99\%$ of its maximum at the optimal timing according to the broadcast mode analysis ($tu_\infty/L_c = 1.35$ for case A and $t/T_p = 0.21$ for case B).  We also quantify the actuation input using the momentum coefficient $C_{\mu}$, defined as,
\begin{equation}
    C_{\mu} = \frac{\iint \rho {u}^{2}_{\text{act}} (\boldsymbol{x}_s,~t) {\rm d}\boldsymbol{x}_s {\rm d}t}{\rho_{\infty} u^{2}_{\infty} L_{c} T^*},
\end{equation}
where $T^*$ is the time scale of the wing motion. For case A (pitching flat plate), $T^*$ is chosen to be the ramp-up time interval such that $T^*u_\infty/L_c = 1$.  For case B (plunging SD7003), $T^*$ is chosen to be the plunging period of the motion, i.e.~ $T^* = 0.5T_p$. The momentum coefficients for both cases are also reported in table \ref{tab:kd}, whose values are chosen in accordance to those reported in previous studies \citep{chen2014experimental, zigunov2022bluff, lin2023flow, de2025duty}.

\begin{table}
\vspace{-0.25in}
\begin{center}
\[
	\text{Actuator model:}\quad u_\text{act} (\boldsymbol{x}_s, t) 
		= u_\text{suction} 
		\times 
		\underbrace{
		\exp\left(-\frac{\left|\boldsymbol{x}_s - \boldsymbol{x}_\text{act}\right|^2}{\sigma^2} \right)}_{g(\boldsymbol{x}_s)}
		\times
		\underbrace{\left[
		\frac{e^{\tilde{\nu}(t - t_{\text{on}})}}{1 + e^{\tilde{\nu}(t - t_{\text{on}})}} - \frac{e^{\tilde{\nu}(t - t_{\text{off}})}}{1 + e^{\tilde{\nu}(t - t_{\text{off}})}} 
		\right]}_{\phi(t)}
\]
    \begin{tabular}{cccc}
      Parameters  & pitching flat plate & plunging SD7003  \\[3pt]
      \hline\\[-4pt]
       $t_{\text{on}}$            & $1.29~L_c/u_\infty$ & $0.16~T_p$\\
       $t_{\text{off}}$           & $1.84~L_c/u_\infty$ & $0.39~T_p$\\
       $\tilde{\nu}$              & $80~L_c/u_\infty$    & $10~T_p$\\
       $x_{0}$                    & $0.02~L_{c}$   & $0.04~L_{c}$\\
       $y_{0}$                    & $0.0115~L_{c}$ & $0.029~L_{c}$\\
       $\sigma$                         & $0.012~L_{c}$  & $0.012~L_{c}$\\
       $T^*$         & $1L_c/u_\infty$    & $1~T_p$ \\ 
       $C_\mu$							& $5.5\%$ & $0.04 \%$
    \end{tabular}
    \caption{The velocity boundary condition for the actuator in the simulation setup. The analytical forms and parameters for the temporal and spatial profiles of actuation are provided here.}
    \label{tab:kd}
\end{center}
\end{table}

\begin{figure}
    \centering
    \begin{overpic}[scale = 1, trim = {0cm, 0cm, 0cm, 0cm}, clip]{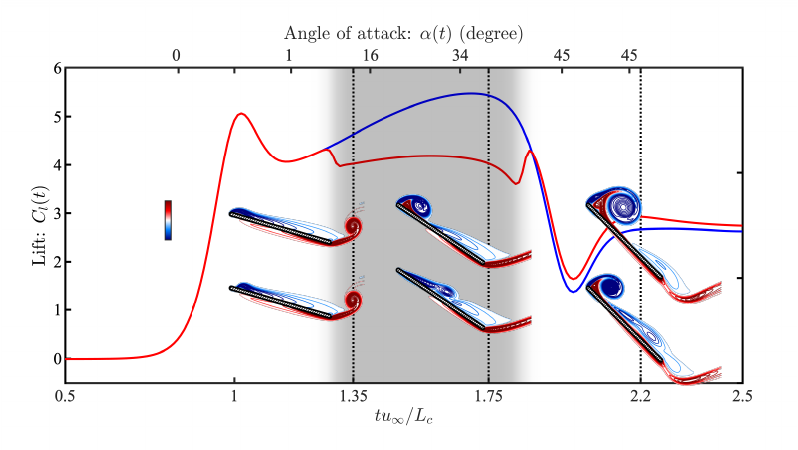}        
        \put(03, 54){\small (a) case A: pitching flat plate}
    
        \put(46, 44){\indexsize \textcolor{blue}{Baseline}}
        \put(50, 38){\indexsize \textcolor{red}{Controlled}}
    
        \put(28, 31.5){\indexsize \textcolor{blue}{Baseline}}
        \put(28, 22){\indexsize \textcolor{red}{Ctrl'd}}
    
        \put(47, 33){\indexsize Control {\bf ON}}
        
        \put(30, 14){\indexsize {(i)}}
        \put(55, 29){\indexsize {(ii)}}
        \put(74, 34){\indexsize {(iii)}}
        
        \put(16, 32){\scriptsize $\omega_{z}L_{c}/u_{\infty}$}
        \put(18, 30){\scriptsize $20$}
        \put(16.5, 26){\scriptsize $-20$}        
    \end{overpic}

    \begin{overpic}[scale = 1, trim = {0cm, 0cm, 0cm, 0cm}, clip]{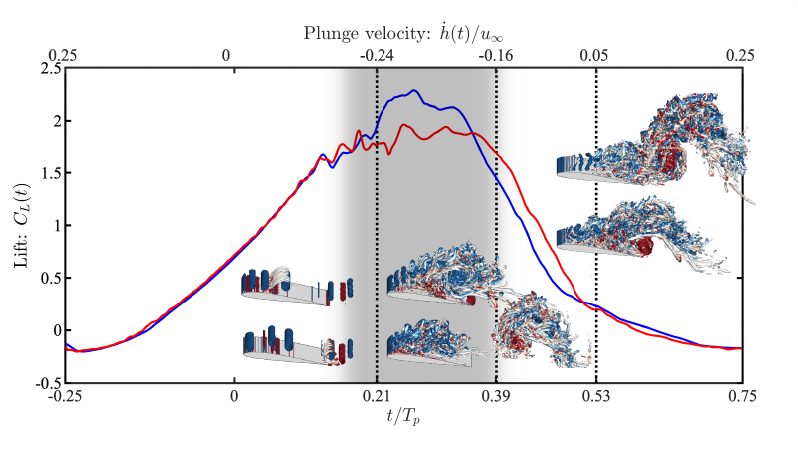} 
        \put(03, 54){\small (b) case B: plunging SD7003}

        \put(54, 45.5){\indexsize \textcolor{blue}{Baseline}}
        \put(50, 34.5){\indexsize \textcolor{red}{Controlled}}
    
        \put(26, 17.0){\indexsize \textcolor{blue}{Baseline}}
        \put(26, 09.5){\indexsize \textcolor{red}{Ctrl'd}}
            
        \put(32, 24){\indexsize {(i)}}
        \put(49, 24){\indexsize {(ii)}}
        \put(70, 38){\indexsize {(iii)}}    
    \end{overpic}
    
    \caption{Comparison of lift fluctuations between the baseline and controlled flows for (a) case A and (b) case B. The spanwise vorticity fields for case A and $Q$-isosurface colored by spanwise vorticity for case B are provided at three representative time instants for both baseline and controlled flows. For case B, the lift coefficient, $C_{L}(t)$, for both the baseline and control cases is cycle-averaged over 4 periods of heaving-plunging motion.}
    \label{fig:cl_control_FP}
\end{figure}
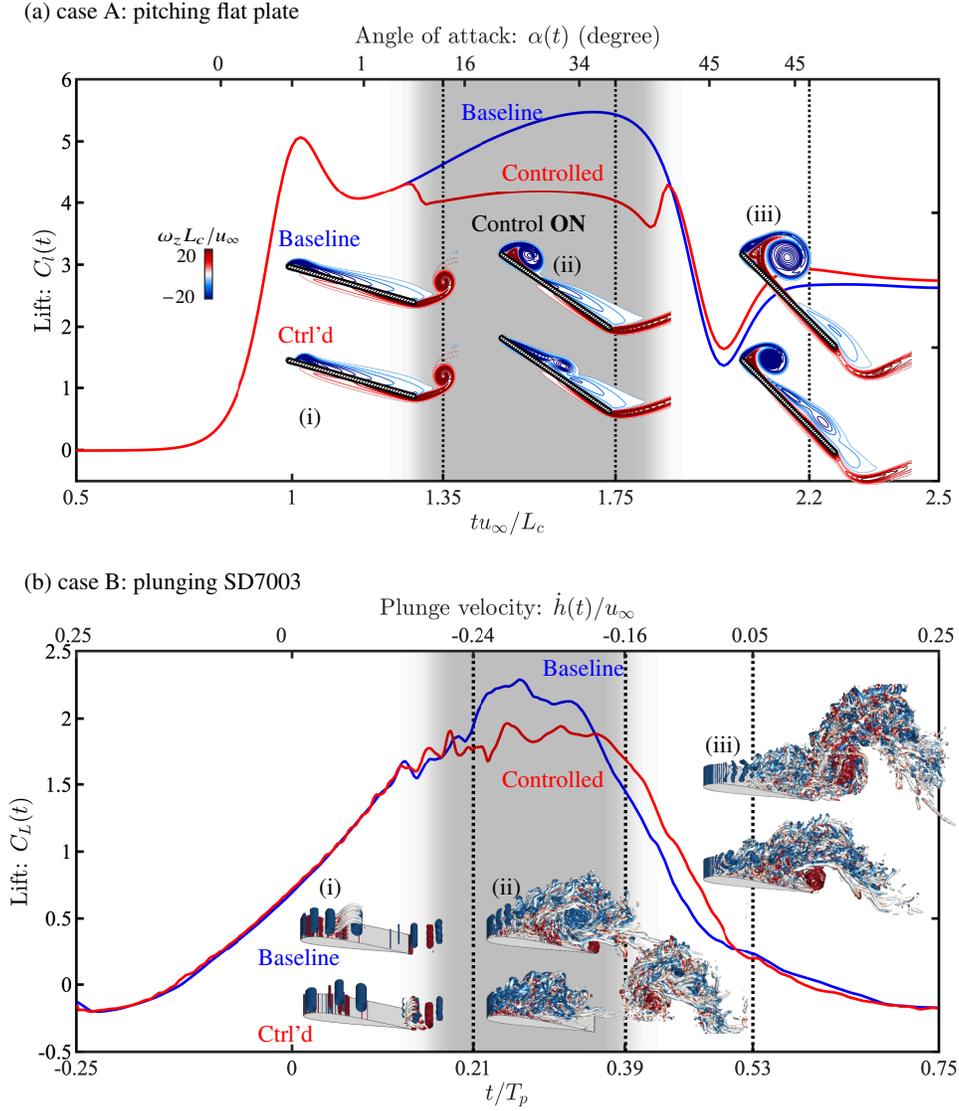

The results of broadcast mode analysis guided flow control for both cases are shown in figure~\ref{fig:cl_control_FP}.  For both cases, the lift fluctuations in time for baseline and controlled flows are provided with instantaneous flow visualizations at three representative instants.  Note that the temporal profiles of control inputs are shown by the shaded regions for both cases. The instantaneous angle of attack and plunge velocity are also shown along to the horizontal axis on the top. For both cases, we observe significant decrease in lift fluctuation. For case A (pitching flat plate), a $21\%$ reduction in peak lift is observed 
as shown in figure~\ref{fig:cl_control_FP}a. More excitingly, we observe $14\%$ decrease in peak lift for the 3D turbulent flow over the plunging SD7003 airfoil, 
as shown in figure~\ref{fig:cl_control_FP}b. These reductions in lift fluctuation for both cases suggests successful mitigation of dynamic stall using the broadcast mode analysis guided flow control.   

In addition to the successful reduction of lift fluctuations achieved by the analysis guided flow control, the flow visualizations inserted in figure~\ref{fig:cl_control_FP} also show that the physical mechanism that results in these reductions agree with the ideation of the broadcast mode analysis and the insights of receiving modes. Flow modifications can be observed starting from instants (i) for both cases, which mark the optimal timings indicated by the broadcast mode analysis. At instants (i) for both cases, we observe that the flow physics near the leading edge are significantly disrupted by the control inputs. This is particularly clear in the visualizations for case A, where we observe that the leading-edge separation at instant (i) ($tu_\infty u/L_c = 1.35$) is effectively suppressed in the the controlled flow. This critical modification interferes with the formation process of the DSV core. This is evident at instant (ii) ($tu_\infty u/L_c = 1.75$) in figure~\ref{fig:cl_control_FP}a. At this instant, a DSV has formed near the LE in the baseline flow.  However, in the controlled flow we observe that the vorticity accumulated near the LE at instant (i) continues to convect along the suction surface and forms a small vortex near the mid-chord at instant (ii). Moreover, this small vortex does not develop into the DSV observed at instant (iii). Instead, the DSV at instant (iii) is developed by accumulating the vorticity in between the LE and the small vortex at instant (ii). These observations show that the control input interferes with the DSV formation by dividing the accumulated vorticity into two vortical structures. Notably, instant (ii) is also the instant where the lift reaches its maximum value. For instant (iii) at $tu_{\infty}/L_{c} = 2.2$, the formations of DSV are concluded for both baseline and controlled flows. At this instant, we observe significantly smaller DSV in the controlled flow, suggesting that the reduction in lift fluctuation is indeed achieved by the modification of the formation process of the DSV.

Similar observations can be made for case B (plunging SD7003) presented in figure~\ref{fig:cl_control_FP}b. For this high Reynolds number 3D turbulent flow, the suction control disrupts both the roll-up and vortex merging processes, as indicated at instant (i) ($t/T_p = 0.21$).  In the baseline flow, the spanwise vortical structures developed from the shear layer roll-up coalesce and form the DSV core. In the controlled flow, the shear layer rolls into smaller vortical structures over the suction surface. Moreover, we observe that the control inputs discourage the spanwise rollers from immediately merging and forming the DSV core. Instead, these spanwise vortices continue to convect downstream past the trailing edge. As a consequence, the formation of the DSV core is postponed due to the control input, similar to what we observed in case A. In particular, the DSV in the control flow observed at instant (ii) ($t/T_p = 0.39$) forms from the breakdown of the leading edge flow into turbulent structures, bypassing the merging process between laminar spanwise rollers in the baseline flow. The DSV at instant (ii) also shows a smaller size in the controlled flow, compared to that in the baseline. The strong DSV in the baseline flow triggers formation of a coherent trailing edge vortex of positive vorticity, which is not observed in the controlled flow until instant (iii). At instant (iii), we observe the departure of DSV from the airfoil surface in the baseline flow. Following the DSV departure, gradual flow reattachment is observed during the upstroke motion. For the controlled flow, the higher lift after $t/T_p = 0.39$ than the baseline flow can be attributed to the delayed DSV departure due to its delayed formation. 

\begin{figure}
    \vspace{0.2in}
    \centering
    \begin{overpic}[scale = 1, trim = {0cm, 0cm, 0cm, 0cm}, clip]{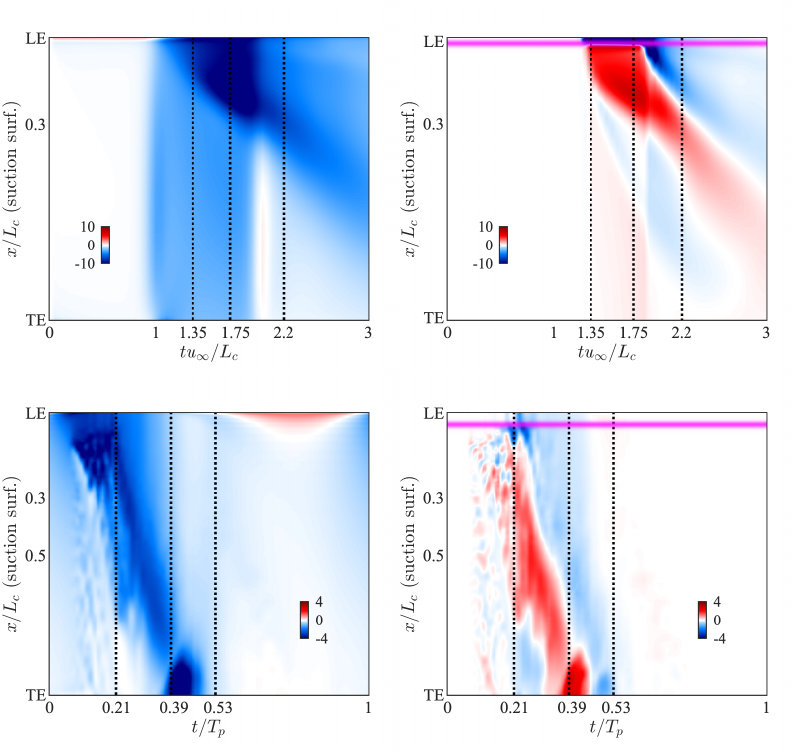}        

    \put(00, 93){\small Case A: pitching flat plate}
    \put(00, 90){\indexsize (a)}
    \put(50, 90){\indexsize (b)}
    \put(00, 46){\small Case B: plunging SD7003}
    \put(00, 43){\indexsize (c)}
    \put(50, 43){\indexsize (d)}
    
    \put(57, 86){\indexsize \textcolor{magenta}{Actuator region}}
    \put(78, 39){\indexsize \textcolor{magenta}{Actuator region}}

    \put(22, 90){\indexsize (i)}
    \put(26, 90){\indexsize (ii)}
    \put(34, 90){\indexsize (iii)}

    \put(72, 90){\indexsize (i)}
    \put(78, 90){\indexsize (ii)}
    \put(83, 90){\indexsize (iii)}
    
    \put(13, 43){\indexsize (i)}
    \put(20, 43){\indexsize (ii)}
    \put(26, 43){\indexsize (iii)}

    \put(64, 43){\indexsize (i)}
    \put(70, 43){\indexsize (ii)}
    \put(77, 43){\indexsize (iii)}

    \put(36, 11){\indexsize $C_{p}(x)$}
    \put(85, 11){\indexsize $\Delta C_{p}(x)$}

    \put(10, 59){\indexsize $C_{p}(x)$}
    \put(60, 59){\indexsize $\Delta C_{p}(x)$}
    
    \end{overpic}

    \caption{The evolution of suction-surface pressure distribution, $C_{p}(x) = (p(x) - p_{\infty})/0.5 \rho_{\infty} u^{2}_{\infty}$, for the baseline flows and the difference in the distribution between the baseline and controlled flows for case A (panels a and b) and case B (panels c and d) is presented.  Representative time instants indicated in figure~\ref{fig:cl_control_FP} are marked along the horizontal axis in each panel.  Note that the actuator over the suction surface is highlighted by a shaded magenta region for both cases in (b) and (d).}
    \label{fig:cp_control_FP}
\end{figure}
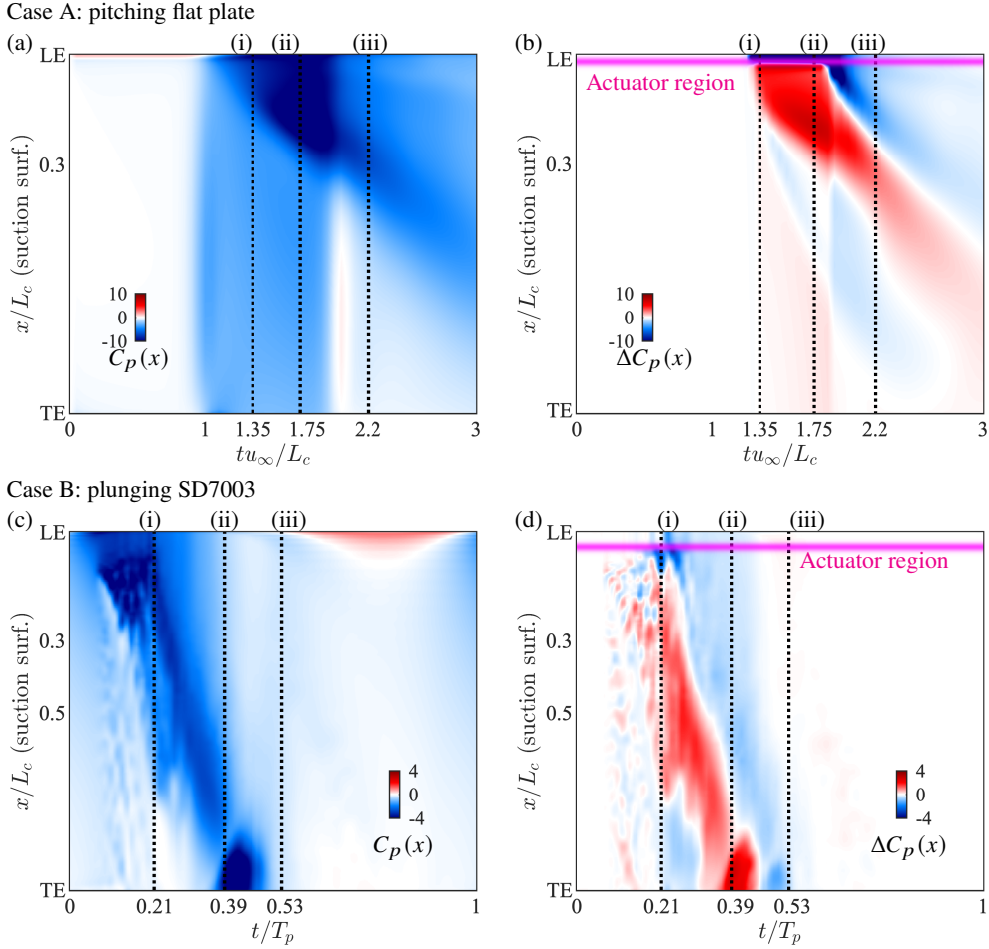

The modified DSV dynamics leaves a clear signature in the suction-surface pressure distribution, which leads to the change in lift fluctuation. This is shown in figure~\ref{fig:cp_control_FP}, where the temporal evolution of the suction-surface pressure distribution ($C_{p}$) for the baseline flows and the change in the suction-surface pressure distribution ($\Delta C_{p}$) for the controlled flows are shown for both cases. For case A, the baseline flow at $tu_{\infty}/L_{c} = 1.35$, marked as (i) in figure~\ref{fig:cp_control_FP}a, exhibits a suction peak near the leading edge. This is the same time instant corresponding to the shear layer pinch-up instant, also marked by (i) in figure~\ref{fig:cl_control_FP}a. This suction results from the accelerated flow around the leading edge directed from the stagnation point on the bottom surface of the airfoil. The suction region progressively expands over the suction surface, particularly in the region $x/L_{c} < 0.3$ during the interval $tu_{\infty}/L_{c} < 2$, coinciding with the DSV formation above the suction surface. This suction induced by the formation of DSV is modified in the controlled flow, as indicated in figure~\ref{fig:cp_control_FP}b. Downstream of the actuator region, a positive $\Delta C_{p}$ indicates the pressure recovery, which leads to the reduction in lift. This recovery is related to both the delayed formation of DSV and also its weaker strength.  

We observe similar patterns for suction-surface pressure distribution for case B, as shown in figure~\ref{fig:cp_control_FP}c. Between $tu_{\infty}/L_{c} = 0$ and $0.21$ (instant (i)), the spanwise vortices from shear layer roll-up, as observed in figure~\ref{fig:cl_flow}b, give rise to the low amplitude pressure fluctuations traveling downstream, as shown in figure~\ref{fig:cp_control_FP}c. These rollers merge at instant (i) and form the core of DSV, which builds up strong suction over the region of $x/L_c < 0.3$.  After this instant, the DSV convects over the suction surface at a lower speed than the spanwise rollers observed in $tu_{\infty}/L_{c} = [0, 0.21]$, as indicated by the dark blue region in figure~\ref{fig:cp_control_FP}c. Under the influence of control, we again observe pressure recovery starting from instant (i), due to the delayed formation of DSV. Moreover, the reduced suction over the region $x/L_{c} > 0.5$ during $t/T_{p} \in [0.21, 0.39]$ reflects the diminished impact of DSV in the controlled flow compared to baseline. Note here that for $t/T_{p} \in [0.21, 0.39]$, there is an enhanced suction for $x/L_{c} < 0.5$ which is due to the delayed presence of DSV. However, the overall change in pressure distribution is favorable in resulting in reduced lift fluctuation observed in figure~\ref{fig:cl_control_FP}.

\begin{figure}
    \centering
    \begin{overpic}[scale = 1, trim = {0cm, 0cm, 0cm, 0cm}, clip]{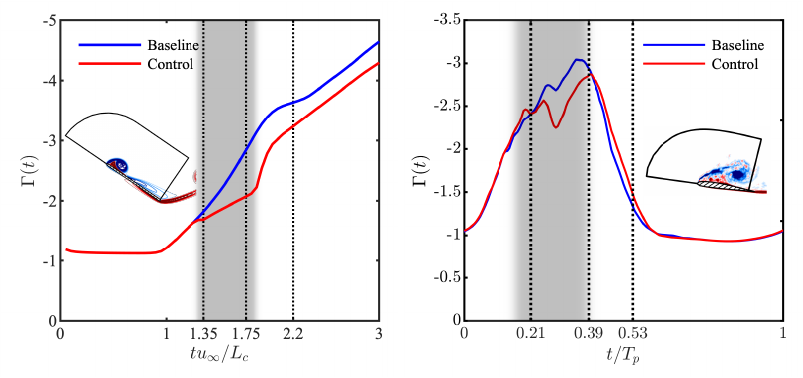}        

    \put(6, 47){\indexsize (a) Case A: pitching flat plate}
    \put(57, 47){\indexsize (b) Case B: plunging SD7003}

    \put(12, 35){\indexsize Bounding box}
    \put(82, 32){\indexsize Bounding box}

    \put(24, 45){\indexsize (i)}
    \put(30, 45){\indexsize (ii)}
    \put(35, 45){\indexsize (iii)}

    \put(65, 45){\indexsize (i)}
    \put(71, 45){\indexsize (ii)}
    \put(77, 45){\indexsize (iii)}
    
    \end{overpic}

    \caption{The temporal evolution of negative vorticity within the bounding box for both the baseline and control flows in cases A and B. The bounding boxes are chosen such that the entire suction surfaces of the airfoils are covered.}
    \label{fig:bound_box_vorticity}
\end{figure}

Recall that the objective function of the broadcast mode analysis considers the reduction in circulation injection during the shedding of LEVs. Here, we analysis the physical flows obtained from the LES to show that this objective is indeed achieved using the of the model-informed suction control. Here, we follow \citet{visbal2018analysis} to quantify the vorticity injection that forms the DSV by integrating the negative vorticity within a chosen bounding box via,
\begin{equation} \label{eq:vort_flux}
    \Gamma (t) = \int_{\boldsymbol{x}\in\text{bounding box}} \omega_{z}^*(\boldsymbol{x}, t) {\rm d}v(\boldsymbol{x}),
\end{equation}
where $\omega_{z}^*$ represents negative spanwise vorticity that characterizes the DSV.  A higher negative value of $\Gamma(t)$ indicates a higher levels of negative vorticity introduced into the bounding box, which can attribute to a stronger DSV.

The modifications in $\Gamma (t)$ and the choices of the bounding boxes for both cases are shown in figure \ref{fig:bound_box_vorticity}.  We observe significantly decreased vorticity injection in the controlled flows for both cases.  For the baseline flow in case A, the non-zero vorticity within the bounding box is observed prior to the onset of pitching at $tu_{\infty}/L_{c} < 1$.  This is primarily due to the presence of the boundary layer along the airfoil surface. Following the initiation of pitching at $tu_{\infty}/L_{c} = 1$, the amount of negative vorticity within the bounding box increases due to accumulation of vorticity within the bounding box to form the DSV core. This is similar to the observations in figure \ref{fig:cl_control_FP}a, where the shear layer continuously feeds negative vorticity into the DSV. With suction control, a reduction in $\Gamma(t)$ is observed at $tu_{\infty}/L_{c} = 1.35$  for case A. This is marked as instant (i) in figure \ref{fig:cl_control_FP}a, which is the same instant in figure~\ref{fig:cl_flow}a that marks the initiation of DSV core formation in the baseline flow. The reduction in $\Gamma(t)$ becomes more pronounced at $tu_{\infty}/L_{c} = 1.75$, marked as instant (ii), where the lift reaches its peak in the baseline flow. At this instant, a $26 \%$ reduction in vorticity within the bounding box is observed, suggesting that the reduction in lift fluctuation (shown figure~\ref{fig:cp_control_FP}b) is indeed attributed to a weakened DSV in the controlled flow, agreeing with the objective of the broadcast mode analysis.  

For case B presented by figure~\ref{fig:bound_box_vorticity}b, we observe similar reduction in the vorticity content within the bounding box but of a different control mechanism. For instant (i) at $t/T_{p} = 0.21$, the suction control does not result in notable decrease in $\Gamma(t)$. However, the small spanwise rollers in the controlled flow continue to travel along the suction surface without immediately merging near the leading edge to form the DSV core. This leads to a rapid decrease in $\Gamma(t)$ past $t/T_p = 0.21$. The formation of DSV is delayed in the controlled flow for instant (ii), where we observe significant decrease in $\Gamma(t)$ within the bounding boxover $t/T_{p} \in [0.21, 0.35]$. The weaker DSV formed in the controlled flow suggests lower injection of negative vorticity, which is also consistent with the objective of the broadcast mode analysis of the vortical network. 

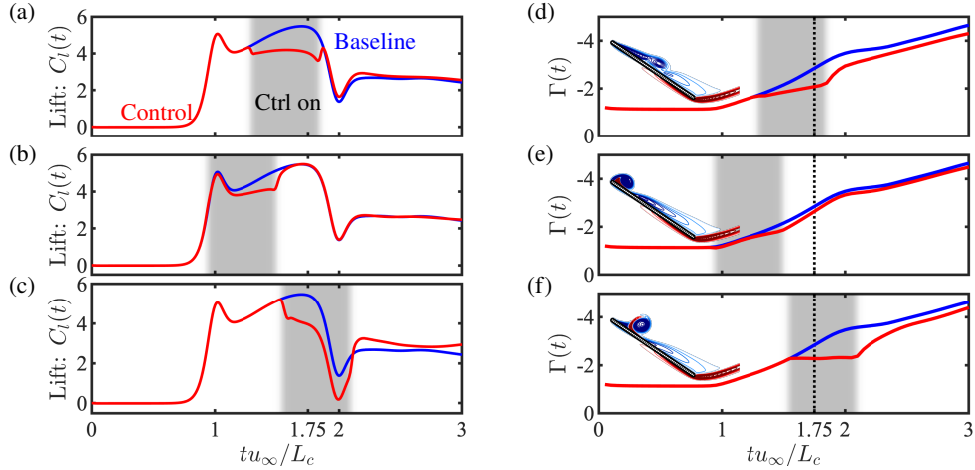
\begin{figure}
    \centering
    \begin{overpic}[scale = 1, trim = {0cm, 0cm, 0cm, 0cm}, clip]{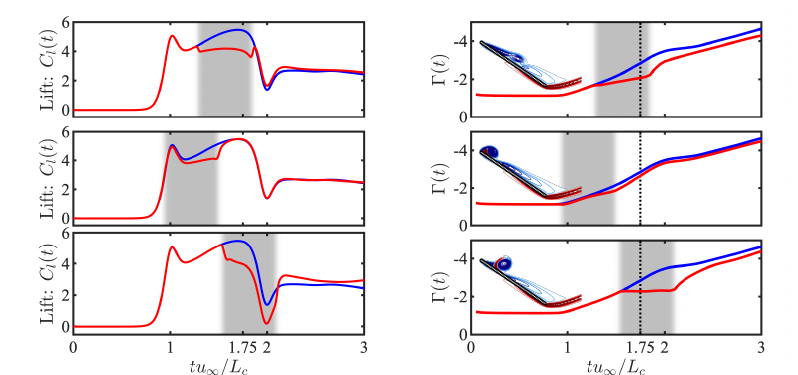}        
        \put(1, 44){\indexsize (a)}
        \put(1, 30){\indexsize (b)}
        \put(1, 17){\indexsize (c)}

        \put(52, 44){\indexsize (d)}
        \put(52, 30){\indexsize (e)}
        \put(52, 17){\indexsize (f)}

        \put(33, 41){\indexsize \textcolor{blue}{Baseline}}
        \put(12, 34){\indexsize \textcolor{red}{Control}}

        \put(25.2, 35){\indexsize Ctrl on}
    \end{overpic}

    \caption{The time histories of the lift coefficient are presented for three actuation cases: (a) actuation at the optimal time, $tu_{\infty}/L_{c} = 1.35$; (b) early actuation at $tu_{\infty}/L_{c} = 1$; and (c) delayed actuation at $tu_{\infty}/L_{c} = 1.5$. The corresponding time histories of negative circulation in the flow are shown in (d), (e) and (f), respectively, along with the spanwise vorticity field at $tu_{\infty}/L_{c} = 1.75$ for control each case.}
    \label{fig:timing_compare}
\end{figure}

Lastly, we consider in figure~\ref{fig:timing_compare} three different timings to turn on the control inputs to highlight the effectiveness of the model-guided flow control. Here, we focus on the case of 2D flow over a pitching flat plate (case A). In addition to the optimal timing guided by the broadcast mode analysis ($tu_{\infty}/L_{c} = 1.35$, figures~\ref{fig:timing_compare}a and \ref{fig:timing_compare}d), LES of controlled flows are also performed with an early actuation at the beginning of pitching ($tu_{\infty}/L_{c} = 1$, figures~\ref{fig:timing_compare}b and \ref{fig:timing_compare}e) and a delayed actuation when the airfoil is midway through its pitching motion ($tu_{\infty}/L_{c} = 1.5$, figures~\ref{fig:timing_compare}c and \ref{fig:timing_compare}f). For these three cases, we examine the control effects by comparing their lift fluctuations (figures~\ref{fig:timing_compare}a-c) and the circulations in the bounding box (figures~\ref{fig:timing_compare}d-f) to those in the baseline flow. Flow visualizations using vorticity field at the instant $tu_{\infty}/L_{c} = 1.75$ where the baseline flow exhibits peak lift are also inserted in figures~\ref{fig:timing_compare}d-f for comparison.

Compared to the case with optimal timing for actuation, the cases with both early and delayed actuation exhibit suboptimal control effects with respect to the reduction of lift fluctuation. For the case with early actuation, the lift recovers to the same level of that in the baseline flow after the control input is turned off. This case exhibits lowest reduction of circulation in the bounding box among the three cases at $tu_{\infty}/L_{c} = 1.75$ and fails to interfere with the formation of DSV and reduce peak lift. These observations are consistent with the insights from the broadcast mode analysis, which indicates that perturbing the vortices prior to optimal time of $tu_{\infty}/L_{c} = 1.35$ has minimal effect on reducing the LEV strengths and, consequently, the development of DSV. On the other hand, for the case with delayed actuation the initial formation of DSV was not interfered and the reduction in peak lift. Instead, the delayed control input prevents the shear layer from continuously feeding vorticity into the DSV, leading to the earlier departure of DSV compared to the other two cases. Comparing this case with the optimal one, although the same levels of lift reduction are observed at $tu_{\infty}/L_{c} = 1.75$, the lift increase in the early stage of DSV formation is not circumvented and results in suboptimal control effectiveness in terms of peak lift. Overall, we observe that the case with optimal timing not only achieves high reduction in peak lift, but the circulation in the bounding box at $tu_{\infty}/L_{c} = 3$ as the DSV convects away is also the lowest among the three cases. This highlights the use of broadcast mode analysis for guiding flow control of dynamic stall flows.

The broadcast mode analysis was designed to reduce the strength of leading edge vortices.  Its results motivate us to target the flow near the leading edge at the critical instants associated with the shear layer pinch-up.  The receiving mode revealed that the circulation of leading edge vortices residing in the DSV core at later times can be reduced.  For both cases considered, a reduction in peak lift is observed in the LES of controlled flows, as shown in figure~\ref{fig:cl_control_FP}. This results from the reduction in the suction induced by the DSV as in figure~\ref{fig:cp_control_FP}, attributed to the formation of weaker vortex core from reduced injection of negative vorticity, as confirmed by figure~\ref{fig:bound_box_vorticity}. These observations are in strong agreement with predictions from the network based analysis of discrete vortices.

\section{Conclusion} \label{sec:conclusion}

We performed broadcast mode analysis of the vortical networks arising from dynamic stall flows to identify the optimal actuation location, timing, and direction for suppressing the dynamic stall. This approach is demonstrated on two flows: a two-dimensional flow over a flat plate airfoil at $Re = 1,000$, and a heaving-plunging spanwise-periodic three-dimensional turbulent flow over a SD$7003$ airfoil at $Re = 60,000$.  The framework leverages the time varying free-vortex states obtained from the discrete vortex method (DVM) to understand the evolution of displacement perturbations to influence the circulation strength of leading edge vortices within the vortical network.  The broadcast mode analysis is reformulated as an optimization problem that seeks the effective vortical nodes and the optimal timing in reducing the circulation strength of leading edge vortices (LEVs), with the objective of mitigating dynamic stall.  The results of the analysis showed that the vortical nodes along the shear layer are the most effective nodes that appear at the instant of shear layer pinch-up. The analysis also showed that the broadcast mode based perturbation obtained from the optimal timing propagates to the vortical nodes residing in the DSV core at a later time. These insights from the broadcast mode analysis are used to inform the choice of actuator location, timing and direction in the simulation. 

To validate the effectiveness of the network-guided actuation strategy, we implemented an actuator positioned near the leading edge in the large-eddy simulations (LES) and activated it at the optimal timing in the direction suggested by the network analysis. Simulation results show a significant reduction in peak lift by $21 \%$ for the 2D flows over the pitching flat plate airfoil and $14 \%$ for the 3D turbulent flow over the plunging SD7003 airfoil. The effect of flow control is found to modulate the negative vorticity injection into the dynamic stall vortex, consistent with the prediction of the broadcast mode analysis that the vortices within DSV core is affect by vortical perturbations.

The observed reduction in the peak lift for both cases underscores the effectiveness of the network-based approach for determining actuator location, timing, and direction for dynamic stall mitigation by targeting the vorticity injection into the DSV. This study establishes a robust framework for active flow control design in the context of unsteady and transient flow separation, highlighting the potential of network-based analysis for flow control design.

\section{Acknowledgments}
We acknowledge the computing resources provided by North Carolina State University High Performance Computing Services Core Facility. The 3D turbulent flow simulations were performed using computational resources sponsored by the Department of Energy's Office of Energy Efficiency and Renewable Energy and located at the National Renewable Energy Laboratory.

\section{Funding}
We gratefully acknowledge the Army Research Office (W911NF-23-1-0109) for supporting this work.

\section{Declaration of interests}
The authors report no conflict of interest.

\bibliographystyle{abbrvnat}

\end{document}